\documentclass[
twocolumn,
pre,
aps,
superscriptaddress,
showpacs,
preprintnumbers,
amsmath,
amssymb]{revtex4-1}

\usepackage{graphicx, subfigure}
\usepackage{amsfonts,bbm}
\usepackage{mathrsfs}
\usepackage{url}
\usepackage{hyperref}% add hypertext capabilities
\graphicspath{{./imgCTRW/}} 

\usepackage{tikz}
\usetikzlibrary{positioning}

\usepackage{color}

\usepackage[framemethod=tikz]{mdframed}
{\begin{mdframed}[hidealllines=true,backgroundcolor=blue!20]}
{\end{mdframed}}

{\begin{mdframed}[hidealllines=true,backgroundcolor=red!20]}
	{\end{mdframed}}

\usepackage{amsthm}

\theoremstyle{plain}
\newtheorem{thm}{Theorem}

\theoremstyle{definition}

\theoremstyle{remark}
\newtheorem{rmk}{Remark}

\begin{document}
	\title{Random walk on temporal networks with lasting edges}
	
	\author{Julien Petit}
	\email{julien.petit@unamur.be}
	\affiliation{Mathematics Department, Royal Military Academy, Brussels (Belgium)}%Lines break automatically or can be forced with \\
	\affiliation{naXys, Namur Institute for Complex Systems, Namur (Belgium)}%
	
	\author{Martin Gueuning}%
	\affiliation{naXys, Namur Institute for Complex Systems, Namur (Belgium)}%
	\affiliation{ICTEAM, Universit\'e catholique de Louvain, Louvain-la-Neuve (Belgium)}%
	
	\author{Timoteo Carletti}
	\affiliation{naXys, Namur Institute for Complex Systems, Namur (Belgium)}

	\author{Ben Lauwens}%
	\affiliation{Mathematics Department, Royal Military Academy, Brussels (Belgium)}

	\author{Renaud Lambiotte}%
	\affiliation{Mathematical Institute, University of Oxford, Oxford (UK)}%
	
	\date{\today}% It is always \today, today,

\begin{abstract}
We consider random walks on  dynamical networks where  edges appear and disappear during finite time intervals. The process is grounded on three independent stochastic processes determining  the walker's waiting-time, the up-time and down-time of edges activation. We first propose a comprehensive analytical and numerical treatment  on  directed  acyclic graphs. Once cycles are allowed in the network, non-Markovian trajectories may emerge, remarkably even if the walker and the evolution of the network edges are governed by memoryless Poisson processes. We then introduce a general analytical framework to characterize  such non-Markovian walks and validate our findings with numerical simulations. 
\end{abstract}

\maketitle

\section{Introduction}

Random walks play a central role in different fields of science \cite{balescu1997statistical,Benavraham2000book,klafter2011first}. Despite the apparent simplicity of the process, the study of random walks remains an active domain of research \cite{fedotov2010non,angstmann2013continuous,angstmann2013forcing-trapping,angstmann2013pattern,Kutner_2017}. Within the field of network science, a central theme focuses on the relation between  patterns of diffusion and  network structure \cite{masuda2017random}.  Important applications include the design of centrality measures based on the density of walkers on nodes \cite{Brin1998conf}, or community detection methods looking for regions of the network where a walker  remains trapped for long times \cite{Rosvall2008PNAS,Delvenne2010PNAS,Lambiotte2015IEEETransNetwSciEng}. The mathematical properties of random walks on static networks are overall well-established \cite{lovasz1993random}, and essentially equivalent to those of a Markov chain. However, the process becomes much more challenging when the network is itself a dynamical entity, with edges appearing and disappearing in the course of time \cite{HolmeSaramaki2013book_Springer,holme2015modern,masudaB}. The temporal properties of networks have been observed and studied in a variety of empirical systems, and their impact on diffusive processes explored by means of numerical simulations \cite{karsai2011small,Starnini2012random,perra2012random} and analytical tools \cite{delvenne2015diffusion}. 

Mathematical analysis of dynamics on temporal networks often relies on the assumption that links activate during an infinitesimal duration \cite{hoffmann2012generalized}. In the case of random walks,  this framework naturally reduces to standard continuous-time random walk on static, weighted networks. Even in this simplified case, however, the dynamics exhibits interesting properties including the so-called waiting-time paradox. When the dynamics of the edges is Poissonian, trajectories are encoded by a Markov chain, whereas  the timings obtained from a non-Poisson renewal process lead to non-trivial properties such as  the emergence of non-Markovian trajectories.  In that case the trajectory of the walker generally depends on its previous trajectory and not only on its current location 
\cite{Speidel2015PhysRevE,gueuning2017backtracking}. The emergence of non-Markovian trajectories is even more pronounced in situations when the activations of edges are correlated, often requiring the use of higher-order models for the data \cite{scholtes2014causality,lambiotte2018understanding}. However, this whole stream of research neglects an important aspect on the edge dynamics, the non-zero duration of their activation, which has been observed and characterised in a variety of real-life systems, including sensor data 
\cite{gauvin2013activity,zhao2011entropy,Scherrer_2008}. The finite duration of edges availability  has important practical implications, including in community detection \cite{sekara2016fundamental}. Theoretically, some results have been obtained within the framework of switching systems, e.g.  by replacing the constant laplacian matrix $L$ by a time-dependent one $L(t)$ for the diffusion   \cite{stilwell2006sufficient,masuda2013temporal,Petit:2017aa},  but a master equation approach derived from a microscopic model of the dynamics is, to the best of our knowledge, still lacking.

Our main objective  is to develop an analytical framework for random walks on temporal networks with finite  activation times. Given a network of potential connections between a fixed set of nodes, the model is defined by three temporal processes. Each process comes with its own timescale, associated to the motion of the random walker, the duration between two successive activations of the edges and the duration of these activations. In contrast with previous research, we derive a master equation from the model specifications,  without implicitly assuming memoryless dynamics for the walker, and consider  the resulting trajectories of the random  walker  \cite{figueiredo2012characterizing}. The competition between three timescales makes the problem particularly rich and we show how certain  master equations already known in the literature are recovered in limit regimes. 

\begin{figure*}[]
	\centering
	\includegraphics[width=0.99 \textwidth]{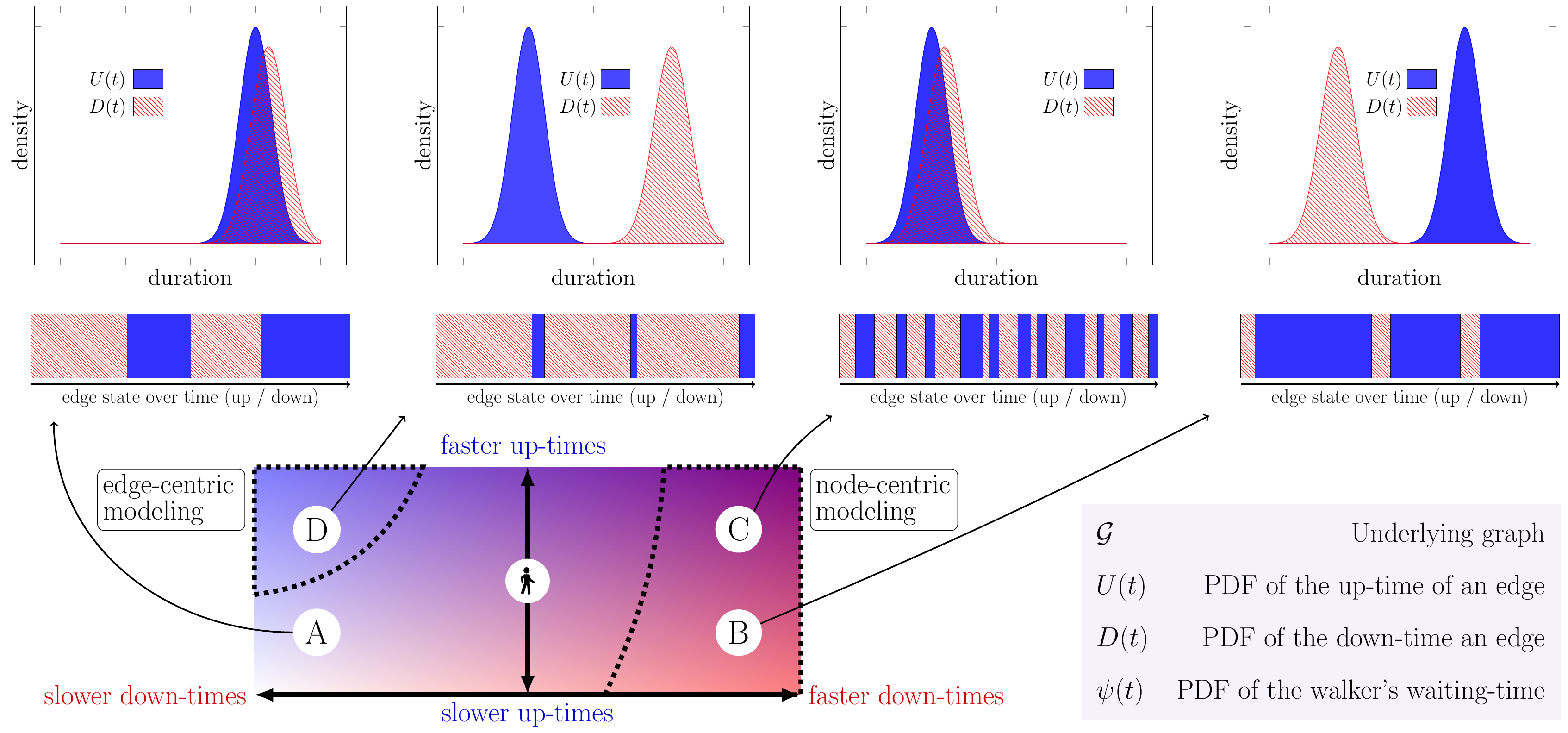}
	\caption{Three timescales are present in the model : one governing the rest state of the walker in the nodes and two associated with the activation and deactivation of the edges. The overall dynamics will depend on the relative weights of such timescales, as we schematically report here.  The bottom panel represents the  up-time and down-time durations of the edges with respect to the walker's self-imposed waiting-time upon arrival on a node. The four corners identified by the letters $A$ to $D$ represent the cases where the timescale for the edges' dynamics is clearly separated from that of  the walker (hereby located in the center of the domain of the bottom panel, and  schematically represented by a standing icon). Each of these four situations is described by one of the four top panels, representing  typical up-time and down-time probability density functions (PDF's)  $U(t)$ and $D(t)$, which complement the walker's possibly node-dependent  PDF $\psi(t)$ as the three dynamical parameters of the model.  The rectangles with the red and blue bars show representative realizations of the stochastic processes of edge activation and deactivation according to the densities reported above.  The blue shaded (resp. red hatched) rectangles stand for periods of edge availability (resp. unavailability). The proposed model depends on the four parameters in the bottom right panel. It is general enough to tackle also  the situations where there is no sharp timescale separation, and none of the dynamics is extreme enough to be neglected, as in the region bounded by the dashed lines in the bottom panel.}
	\label{fig:cartoon}
\end{figure*}

This paper is organised as follows. In section \ref{sec:model}, we describe the model and its  parameters. In section~\ref{sec:reducible-model}, we derive a master equation for the density of the walker valid for directed acyclic graphs (DAGs).  Particular cases for the model parameters and their ensuing dynamics are discussed. These equations   are revisited in section~\ref{sec:emergenceMemory}, where we consider the impact of cycles in the graph on the Markovianity of the process. The analytical predictions are  confronted with numerical simulations throughout this work. Section~\ref{sec:simulation} gives more details about the numerical implementation of our formalism. We finally conclude and give perspectives in section~\ref{sec:applications}.

\section{The model}\label{sec:model}
\begin{figure}[] 
	\hspace*{-1em}
	\begin{tikzpicture}[scale = .8,transform shape]
	%\draw[help lines,line width=1,gray,thin] (-5, -5) grid (15,15);
	
	% AXIS FOR A_ij 
	\begin{scope}[yshift = 1cm]
	\draw[ thick,->] (0,-.3)--(0,0)node[anchor=south east]{0} 
	--(0,1)node[anchor=east]{1}
	-- (0,1.5) node[anchor=south west]{$A_{13}(t)$};
	\draw(-.15,1)--(.15,1);
	\draw[thick,->](-.3,0) -- (10.5,0)node[anchor = north west](t){$t$};
	
	% EVOLUTION OF A_ij
	%\draw[fill] (0,0)circle(.1); 
	\draw[line width = 2.3pt](0,0)--(2,0) (2,1)--(4,1) (4,0)--(8,0) (8,1)--(10,1);
	\draw[fill = white,very thick] (2,0) circle (.1);
	\draw[fill = white,very thick] (4,1) circle (.1);
	\draw[fill = white,very thick] (8,0) circle (.1);
	%\draw[fill = white,very thick] (10,1) circle (.1);
	
	\draw[fill = black,very thick] (2,1) circle (.1);
	\draw[fill = black,very thick] (4,0) circle (.1);
	\draw[fill = black,very thick] (8,1) circle (.1);
	\end{scope}
	
	% AXIS FOR TIME
	\begin{scope}[yshift = -2.4cm]
	\draw[ thick,->] (-.3,0) -- (10.5,0)node[anchor = north west](t){$t$};
	\draw[thick] (0 cm, 2.5 pt) -- (0 cm,-2.5pt) node[anchor=north] {%
		$t_ {m}$};
	\foreach \x in {2,4,...,10}
	\draw[thick] (\x cm, 2.5 pt) -- (\x cm,-2.5pt) node[anchor=north] {%
		$t_ {m+\pgfmathparse{0.5*\x}% Evaluate the expression
			\pgfmathprintnumber[    % Print the result
			fixed,
			fixed zerofill,
			precision=0
			]{\pgfmathresult}}$};
	\end{scope}

	% SERIES OF ADJACENCY MATRICES
	
	% 1/5
	\begin{scope}[xshift = .5cm,yshift = -0.4cm]
	\path[fill = blue!5!white](-.5,-1.5)rectangle (1.5,.4);
	\node (myfirstpic) at (.05,0.6) {\includegraphics[width=.34cm]{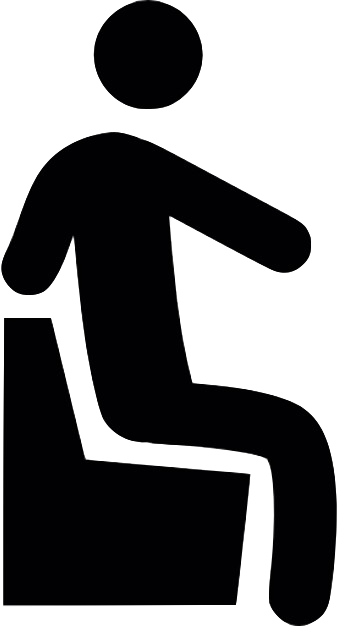}};
	\path[draw,fill = white] 
	(0,0) circle (.2) node[circle,draw](c1){1}
	++(1,0) circle (.2)  node[circle,draw](c2){2}
	++(0,-1) circle(.2) node[circle,draw](c3){3}
	++(-1,0)circle(.2) node[circle,draw](c4){4};
	\draw[line width = 1pt,->] (c1) to[bend left] (c2);
	\draw[line width = 1pt,->] (c4) to[bend right] (c3);
	\node [fill = white] (A) at (.5,-1.9){$\bf G_{m}$};
	\end{scope}
	
	% 2/5
	\begin{scope}[xshift = 2.5cm,yshift = -0.4cm]
	\path[fill = blue!10!white](-.5,-1.5)rectangle (1.5,.4);
	\node (myfirstpic) at (.05,0.6) {\includegraphics[width=.34cm]{sitting.png}};
	\path[draw,fill = white] 
	(0,0) circle (.2) node[circle,draw](c1){1}
	++(1,0) circle (.2)  node[circle,draw](c2){2}
	++(0,-1) circle(.2) node[circle,draw](c3){3}
	++(-1,0)circle(.2) node[circle,draw](c4){4};
	\draw[line width = 1pt,->] (c1) to[bend left] (c2);
	\draw[line width = 1pt,->] (c1) to[bend left] (c3);
	\draw[line width = 1pt,->] (c4) to[bend right] (c3);
	\node [fill = white] (A) at (.5,-1.9){$\bf G_{m+1}$};
	\end{scope}
	
	% 3/5
	\begin{scope}[xshift = 4.5cm,yshift = -0.4cm]
	\path[fill = blue!15!white](-.5,-1.5)rectangle (1.5,.4);
	\node (myfirstpic) at (.05,0.6) {\includegraphics[width=.34cm]{sitting.png}};
	\path[draw,fill = white] 
	(0,0) circle (.2) node[circle,draw](c1){1}
	++(1,0) circle (.2)  node[circle,draw](c2){2}
	++(0,-1) circle(.2) node[circle,draw](c3){3}
	++(-1,0)circle(.2) node[circle,draw](c4){4};
	\draw[line width = 1pt,->] (c1) to[bend left] (c2);
	%\draw[line width = 1pt,->] (c1) to[bend left] (c3);
	\draw[line width = 1pt,->] (c4) to[bend right] (c3);
	\node [fill = white] (A) at (.5,-1.9){$\bf G_{m+2}$};
	\end{scope}
	
	% 4/5
	\begin{scope}[xshift = 6.5cm,yshift = -0.4cm]
	\path[fill = blue!20!white](-.5,-1.5)rectangle (1.5,.4);
	\node (myfirstpic) at (.05,0.65) {\includegraphics[width=.38cm]{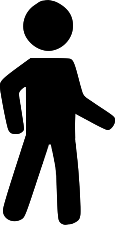}};
	\path[draw,fill = white] 
	(0,0) circle (.2) node[circle,draw](c1){1}
	++(1,0) circle (.2)  node[circle,draw](c2){2}
	++(0,-1) circle(.2) node[circle,draw](c3){3}
	++(-1,0)circle(.2) node[circle,draw](c4){4};
	%\draw[line width = 1pt,->] (c1) to[bend left] (c2);
	%\draw[line width = 1pt,->] (c1) to[bend left] (c3);
	\draw[line width = 1pt,->] (c4) to[bend right] (c3);
	\node [fill = white] (A) at (.5,-1.9){$\bf G_{m+3}$};
	\end{scope}
	
	% 5/5
	\begin{scope}[xshift = 8.5cm,yshift = -0.4cm]
	\path[fill = blue!25!white](-.5,-1.5)rectangle (1.5,.4);
	\node (myfirstpic) at (1.5,-.8) {\includegraphics[width=.85cm]{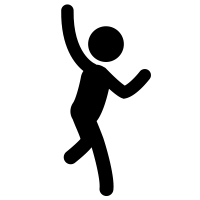}};
	\path[draw,fill = white] 
	(0,0) circle (.2) node[circle,draw](c1){1}
	++(1,0) circle (.2)  node[circle,draw](c2){2}
	++(0,-1) circle(.2) node[circle,draw](c3){3}
	++(-1,0)circle(.2) node[circle,draw](c4){4};
	%\draw[line width = 1pt,->] (c1) to[bend left] (c2);
	\draw[line width = 1pt,->] (c1) to[bend left] (c3);
	\draw[line width = 1pt,->] (c4) to[bend right] (c3);
	\node [fill = white] (A) at (.5,-1.9){$\bf G_{m+4}$};
	\end{scope}
	
	\end{tikzpicture}
	\caption{Directed temporal network with four nodes (below) and the (1,3) entry of the time-dependent adjacency matrix $A(t)$  (above). Note that $A(t)$ is right-continuous and is given by $A(t) = G_i$ for $t_{i}\leq t < t_{i+1}$. In this example, the up-time $t_{m+2}-t_{m+1}$ of edge $1\rightarrow 3$ follows the density  $U_{13}(t)$, while the down-time $t_{m+4}-t_{m+2}$ has density $D_{13}(t)$. At every random rewiring time $t_i$, almost surely only one edge changes.  }
	\label{fig:discrete-switching}
\end{figure}

Let $V$ be a  fixed set of $N$ nodes and $E$ be a set of directed edges between these nodes. 
We denote by $\mathcal G  = (V,E)$ the  static graph determining which edges  are available in the dynamic graph with time-dependent adjacency matrix $A(t)$. The dynamic graph can assume any of the $2^{\left \vert E \right \vert }$ possible configurations allowed by $\mathcal{G}$.  In our model-driven approach, each edge $(i,j) = i\rightarrow j \in E$ is characterized by 
\begin{itemize}
	\item a down-time probability density function (PDF) $D_{ij}(t)$, $t\in \mathbb{R}^+$, which determines for how long the edge remain inactive; 
	\item an up-time PDF $U_{ij}(t)$, $t \in \mathbb{R}^+$, which rules the duration that the edge is available to the walker. 
\end{itemize}
In this work, the random variables associated with the densities $U_{ij}$ and $D_{ij}$ are assumed to have finite expectation. The adjacency matrix can be written as 
\begin{equation}\label{eq:A-of-t}
A(t) = \sum\limits_{i \in \mathbb{Z}}G_i \mathbbm{1}_{ \lbrace t_i \leq t < t_{i+1} \rbrace}(t)
\end{equation}
with $\ldots < t_{-1}<t_0 \leq 0 <t_1 < \ldots$ the successive times of the rewiring, and $G_i$ a fixed adjacency matrix (figure~\ref{fig:discrete-switching}).  Let $k_i^{\rm in}(t)   = \sum_{j = 1}^{N}A_{ji}(t)$ be the in-degree of node $i$ at time $t$, and $k_i^{\rm out}(t)   = \sum_{j = 1}^{N}A_{ij}(t)$ be the out-degree.   We define the set of nodes reachable from $i$ in the underlying graph $V_i  = \left \lbrace j \in V \ | \ i\rightarrow j \in E \right \rbrace$, and  $|V_i|$ its cardinality, namely  the out-degree of node $i$ in $\mathcal{ G}$. Similarly,  $V'_i  = \left \lbrace j \in V \ | \ j\rightarrow i \in E \right \rbrace$  and  $|V'_i| $ results to be the in-degree of node $i$. We make the assumption that there are no isolated nodes in $\mathcal{ G}$ : for every $i \in E$, $\max \lbrace   |V_i|, |V'_i|  \rbrace >0$. 

Let us define the random walk. 
A continuous-time random walk  on a dynamical graph with adjacency matrix $A(t)$, $t\in \mathbb{R}$ is a process $\left \lbrace A(t),{i}_W(t)\right \rbrace$ where ${i}_W(t) \in V$ is the node occupied by the walker at time $t$. Upon arrival on a node $i$, the walker is assigned a waiting-time according to the PDF $\psi_i(t)$ which generally depends on the node (see Fig.~2, first, second and third cartoon from the left). After the waiting-time has elapsed, the walker selects one of the available  leaving edges uniformly, namely with probability $1/k_i^{\rm out}(t)$. If no edge is available, the walker is trapped on the node (Fig.~2, fourth cartoon) and waits for the first leaving edge to appear to perform the jump (Fig.~2, fifth cartoon). Note that in the latter case, almost surely there is no choice to be made there~: no two or more edges can activate at the same time.

Let us observe that a possible variant of this random walk could consist in assigning a new waiting-time according to $\psi_i$ for the walker trapped on a node because of the lack of available edges once it is ready to jump. This process was studied in \cite{figueiredo2012characterizing}, where the authors exclusively focus on the asymptotic state of the process.

Our model is an extension of the standard active node-centric and passive edge-centric random walks in temporal networks \cite{masudaB}, where edge duration is instantaneous. In the former, the motion is determined by the waiting-time of the walker and, once a jump takes place, all the edges in $\mathcal G$ are available - or at least the ones exiting from the node where the walker is located.  In the latter case, the walker is ready to jump as soon as it arrives on a new node, and it takes the first edge that appears - the walker thus passively follows the appearing edges modeled by a renewal process. 
These two cases correspond to asymptotic regimes described by our model when a timescale dominates over the others. In general, however, the process is determined by the competition of three timescales.
Figure~\ref{fig:cartoon} summarizes  possible scenarios  labeled from $A$ to $D$ corresponding to the four distinct cases, where the dynamics of the down- and up-times are either significantly faster or significantly slower than the characteristic waiting-time of the walker. At the right border of the domain, in the region ranging from $B$ to $C$, when the walker is ready to jump the possible extra waiting-time for an edge to become available is usually short, and the network dynamics can be neglected. Therefore, the node-centric random walk is a good proxy for our model. In the region centered around $D$, the same type of analysis leads instead to  neglecting the waiting-time of the walker. In general however, as in the center of the domain, in the area between the dotted regions, neither the walker nor the edges dynamics can be neglected. This region is the focus of our work.

% TABLE WITH MODEL PARAMETERS - NOW INCLUDED IN FIRST FIGURE
%\begin{table}[]%The best place to locate the table environment is directly after its first reference in text
%	\caption{\label{tab:modelparameters}%
%		Model parameters. The adjacency matrix of the temporal graph $A(t)$ results from the model parameters and is given by \eqref{eq:A-of-t}. Note that this modeling differs from a data-driven approach, where $A(t)$ is actually deterministic. Due to the independent densities on the nodes and on the edges, this model is a competitive one between the nodes process, and the edge processes. 
%	}
%	\begin{ruledtabular}
%		\begin{tabular}{lr}
%$\mathcal{ G}$ & Underlying graph\\  
%$U_{ij}(t)$& Density of the up-time of edge $i \rightarrow j$  \rule{0pt}{3.5ex}\\ 
%$D_{ij}(t)$& Density of the down-time of edge $i \rightarrow j$  \rule{0pt}{3.5ex}\\ 
%$\psi_i(t)$\rule{0pt}{3.5ex}& Density of the waiting-time of the walker on node $i$ \\ 
%		\end{tabular}
%	\end{ruledtabular}
%\end{table}

% FIGURE : BEHAVIOR OF TEMPORAL NETWORK

\section{The case of  directed acyclic graphs}\label{sec:reducible-model}
\begin{figure} % condensation
\centering
	\begin{tikzpicture}[scale = .8, transform shape]
	% before condensation
	\draw[help lines] (0,0) grid (4,2);
	\node[draw, style = {circle},fill = blue!25!white](1) at (0,0) {1};
	\node[draw, style = {circle},fill = blue!25!white](2) at (0,2) {2};
	\node[draw, style = {circle},fill = blue!25!white](3) at (2,2) {3};
	\node[draw, style = {circle,fill = yellow!10!white}](4) at (4,2) {4};
	\node[draw, style = {circle},fill = green!75!blue!10](5) at (4,0) {5};
	\node[draw, style = {circle},fill = green!75!blue!10](6) at (2,0) {6};
	\draw[line width = 1.5pt,->] (1)--(2);
	\draw[line width = 1.5pt,->] (2)--(3);
	\draw[line width = 1.5pt,->] (3)--(1);
	\draw[line width = 1.5pt,->] (3)--(4);
	\draw[line width = 1.5pt,->] (4)--(5);
	\draw[line width = 1.5pt,->] (5) to[bend right] (6); 
	\draw[line width = 1.5pt,->] (6) to[bend right] (5); 
	% result of condensation : 
		\begin{scope}[yshift = -1.5cm]
		\node[draw, style = {circle},fill = blue!25!white](s1) at (0,0) {1/2/3};
		\node[draw, style = {circle},fill = yellow!10!white](s2) at (2,0) {4};
		\node[draw, style = {circle},fill = green!75!blue!10](s3) at (4,0) {5/6}; 
		\draw[->,line width = 1.5pt](s1)--(s2);
		\draw[->,line width = 1.5pt](s2)--(s3);
		\end{scope}
	\end{tikzpicture}
	
	\caption{Mapping of a directed network with cycles (top) to a DAG (bottom) through a condensation process. Strongly connected components are transformed into super-nodes. \label{fig:condensation}}
\end{figure}
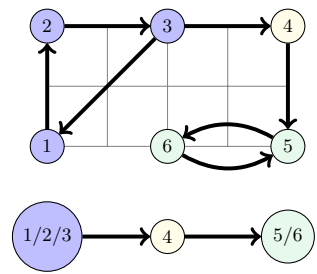

As a first step, we  consider the trajectory of a walker performing a random walk as defined above on a directed acyclic graph (DAG). The reason for that is twofold. 
\begin{enumerate}
	\item DAGs include  directed trees and find many applications, see for instance \cite{melnik2011unreasonable}. Every undirected graph possesses an acyclic orientation. Moreover, by contracting each strongly connected component, every directed graph can be mapped to a DAG. Figure~\ref{fig:condensation} illustrates that process. The material presented in this section   therefore provides tools to analyse a random walk on a coarse grained model obtained by condensation of a given  graph into a DAG. 
	\item As we will show next, the presence of cycles in the graph  will remove the Markov property from the random walk. Hence, the analysis of our model on a DAG will serve, in a second step, as a limiting case on which to consider more general organizations.  The  approximation using DAGs is expected to be good  when edges along a path can be considered statistically independent. The conditions for this to hold  will be discussed further in  section \ref{sec:emergenceMemory}. 

\end{enumerate}
As will become clear (see \ref{sec:modelreduction}), the model on DAGs can be viewed as a one-density, node-centric (or edge-centric) random walk. 

\subsection{The master equation on a DAG} \label{sec:master-eq-DAG}
The notations in this section are adapted from \cite{hoffmann2012generalized}.   Let $n_i(t)$ be the probability for the walker to be on node $i$ at time $t$, 
\begin{equation}\label{eq:ni}
n_i(t) = \rm  P \lbrace i_W(t) = i\rbrace. 
\end{equation}
If $q_i(t)$ is the PDF of the arrival time on node $i$,  and $\Phi_i(t,\tau)$ is to probability to stay on node $i$ on the interval $[\tau,t]$ with $\tau$ the arrival time on node $i$, then 
\begin{equation}\label{eq:ni-qi-Phii}
n_i(t) = \int_0^t q_i(\tau) \Phi_i(t,\tau) \rm d \tau . 
\end{equation}
Let $T_{ji} (t,\tau)$ be the PDF of the transition time from node~$i$ to  $j$, with $\tau$ the arrival time on node $i$. Let also $T_{\bullet i}(t,\tau)$ denote the PDF of the time of the jump from node $i$. We have 
\begin{align}
\Phi_i(t,\tau) & = 1-\int_{\tau }^{t} T_{\bullet i} (\nu , \tau) \rm d \nu  \nonumber \\
					 & = 1-\int_{\tau}^{t} \sum_{j \in V_i} T_{ji}(\nu,\tau) \rm d \nu.  \label{eq:Phi-version2}
\end{align}
We want to write the column vector  ${\bf n }(t) = (n_1(t), \ldots,n_N(t))^T$ in terms of the transition density $T_{ij}$ and of the initial condition ${\bf n} (0)$. Looking at \eqref{eq:ni-qi-Phii} and \eqref{eq:Phi-version2}, we search for an appropriate expression for $q_i(t)$. Let $q_i^{(k)}(t)$ be the probability to arrive on node~$i$ at time~$t$  in exactly $k \in \mathbb{N}$ jumps. Then we have 
\begin{equation}\label{eq:qi-qk}
q_i(t) = \sum\limits_{k=0}^{\infty}q_i^{(k)} (t) = \sum\limits_{k\geq 1} q_i^{(k)}(t) + q_i^{(0)}(t),
\end{equation} 
with  initial condition $q_i^{(0)}(t) = n_i (0) \delta(t)$. Equivalently, 
\begin{equation}\label{eq:qi-qk1}
q_i(t) = \sum\limits_{k= 0}^{\infty} q_i^{(k+1)}(t) + q_i^{(0)}(t),
\end{equation}
where 
\begin{equation}\label{eq:qi-integral-qj-Tij}
q_i^{(k+1)}(t) = \sum_{j} \int_0^t q_j^{(k)}(\nu) T_{ij}(t,\nu) \rm d \nu . 
\end{equation}
Summing on both sides over $k\geq 0$ and adding $q_i^{(0)}(t)$ yields
\begin{equation}\label{eq:qi-fundamental}
q_i(t) = \sum_{j} \int_0^t q_j(\nu) T_{ij}(t,\nu){\rm d} \nu + q_i^{(0)}(t). 
\end{equation}
In vector form, with $\mathbf{q}(t) = \left ( q_1(t),\ldots,q_N(t)\right )^T$, we have
\begin{equation}\label{eq:volterraII}
\mathbf{q}(t) = \mathcal{T} \mathbf{q}(t) + \mathbf q^{(0)}(t)
\end{equation}
where  $\mathcal{T}$ is the linear integral operator acting on $\mathbf{q} (t)$ defined by : 
\begin{equation}\label{eq:operator-T}
\mathcal{T} \mathbf{q}(t) = \int_0^t T(t,\nu)\mathbf q(\nu) \mathrm d \nu, \quad i = 1,\ldots,N
\end{equation}
where $T(t, \nu)$ is a matrix function with component $(i,j)$ given by $T_{ij}(t ,\nu)$. 
Due to the acyclic nature of the graph and as will become clear after remark~\ref{rmk-acyclic} at the end of section~\ref{sec:transition-density-DAG}, the transition density actually only depends on the duration $t-\tau$.  As a result, equation \eqref{eq:operator-T}   is a convolution and applying a Laplace transform   allows to solve \eqref{eq:volterraII}  for $\mathbf{q}(t)$, as was done in \cite{hoffmann2012generalized}. 

Once $\mathbf{ q}(t)$ is found, we consider equation~\eqref{eq:ni-qi-Phii}, which can be cast under the form
\begin{equation}\label{eq:ni-qi-Phii-operator}
\mathbf n (t) = \mathcal P \mathbf q (t)
\end{equation}
where operator  $\mathcal P $ is diagonal and given by 
\begin{equation}\label{eq:P-operator}
\left ( \mathcal P \mathbf q (t)\right )_i = \int_0^t \Phi_i(t,\tau) q_i(\tau) \mathrm d \tau, \quad i = 1,\dots, N. 
\end{equation}
Observe that this is again a convolution, because $\Phi_i(t,\tau)$ is essentially $\Phi_i(t-\tau)$. The right-hand-side of \eqref{eq:P-operator} can be computed directly in the time-domain, or through a Laplace transform. In the latter case we obtain for each component  $\hat { {n}}_i(s):= \int_0^\infty n_i(t)e^{-st}\mathrm d t$ a product in the Laplace domain, and we ultimately find  $\mathbf n (t) = (n_1(t),\ldots,n_N(t))$ as a function of the initial density $\mathbf n (0)$ by computing the inverse transform of these products.

\subsection {On the use of the Laplace transform in the case of a DAG}
It is not mandatory to use the Laplace transform to solve the integral equations $\mathbf q (t)$ and then get $\mathbf n (t)$. We can proceed directly in the time domain and solve the equation relying on the acyclic nature of the graph. We detail this alternative approach, which does not rely on the convolution structure of the integral equations. 
\begin{rmk}
This method also applies  when we drop the acyclic assumption on $\mathcal{ G}$ in section~\ref{sec:emergenceMemory} and we have to solve equation~\eqref{eq:q_tuple3}. 
\end{rmk}

Let us first recall Neumann's Lemma. 
\begin{thm} Let $\mathcal T $ be a linear bounded operator on a Banach space $X$. If $||\mathcal T||= \sup_{||x|| \leq 1} ||\mathcal Tx|| <1$, then $I-\mathcal T$ is invertible and is given by the Neumann series
	$$(I-\mathcal T)^{-1} = \sum_{k=0}^\infty T^k = I + \mathcal T + \mathcal T^2 + \ldots . $$	
\end{thm}
The theorem is applicable for this convolution-type linear Volterra integral equation with square integrable convolution kernels (see \cite{delves1988computational} theorem 3.7.7 page 77), and equation  \eqref{eq:volterraII} gives
\begin{align}
\mathbf q (t) &= (I-\mathcal T)^{-1} \mathbf q ^{(0)}(t) \nonumber \\ % \label{eq:qi-I-T-1}\\
&= \sum_{k=0}^{\infty} \mathcal T ^k \mathbf q^{(0)} (t) \label{qi:Neumann}\\
&= \sum_{k=0}^{\infty} \mathcal T ^k \delta (t) \mathbf n{(0)} . \label{qi-n0-vector}
\end{align}

If we compute the iterates of $\mathcal T $ acting on $\mathbf q^{(0)}(t)$, we see that the successive terms $\mathcal T ^k \mathbf q^{(0)}(t)$, with $\mathbf q^{(0)}(t) = \mathbf n ^{(0)} \delta (t) $, account for the probability to arrive on a given node at time $t$, starting from the initial condition $\mathbf n ^{(0)}$, in exactly $k$ steps.

\begin{rmk}
	In general, the Neumann series does not offer a practical way for computing $(I - \mathcal T)^{-1}$ since it involves an infinite number of terms. Because we make the assumption that the underlying graph $\mathcal{ G}$ has no cycles, the series can  be cut after $d$ terms, where $d$ is the diameter of the graph.  
\end{rmk}

Based on \eqref{eq:ni-qi-Phii} we can now compute $ \dot{\mathbf{n}} (t)$ in terms of the transition density and of the initial conditions. Applying Leibniz's rule for differentiation under the integral sign, we obtain
\begin{align}
{ \dot{n}_i}(t) &= q_i(t) - \int_0^t q_i(\tau) \sum_{j\in V_i} T_{ji}(t,\tau) \mathrm d \tau \nonumber \\ %\label{eq:ni-qi-version1}\\
&= q_i(t) - \int_0^t q_i(\tau) T_{\bullet i}(t,\tau) \mathrm d \tau\label{eq:ni-qi-version2}. 
\end{align} 
The interpretation is that the  rate of evolution of  $n_i(t)$ is given by a  sum of all arrivals minus the departures, with each departure resulting from a previous arrival at any point in time. 
Let us  define a diagonal integral operator $\mathcal D$ acting on $\mathbf q$  by its $i$-th component : 
\begin{equation}\label{eq:operator-D}
\left (\mathcal D  \mathbf q (t) \right )_i  = \int_0^t T_{\bullet i}(t,\tau) q_i(\tau) \mathrm d \tau. 
\end{equation}
Equation \eqref{eq:ni-qi-version2} can now be written as 
\begin{align}\label{eq:ndot-operator-form}
 \dot{\mathbf{n}} (t) &= (I-\mathcal D) \mathbf q (t) \nonumber \\
&= (I-\mathcal D) \sum_{k=0}^\infty \mathcal T ^k \mathbf q^{(0)}(t)
\end{align}
where we have used \eqref{qi:Neumann} to obtain the second equation. 

\subsection{Transition density on DAGs}\label{sec:transition-density-DAG}
The equation for $\mathbf n (t)$ remains abstract unless we can write $T_{ji}$ explicitly in terms of the model parameters contained in figure~\ref{fig:cartoon}.  For the sake of simplicity and without lack of generality in the reasoning, we  assume that 
all edges share  the same up-time and down-time densities : $U_{ij}(t) =:U(t)$ and $D_{ij}(t) =: D(t)$, for all $i,j = 1,\ldots, N$.

Let $p$ denote the probability that a given edge of $\mathcal G$ is active (up-state) at a random  time. Recall that we assume up-time and down-time durations with finite expectation. It results that 
\begin{equation}\label{eq:p-up}
p = \mathrm P \left \lbrace  \mathrm{edge \ }i\rightarrow j  \ \mbox{is active} \right \rbrace = \frac{\langle U \rangle}{\langle U \rangle + \langle D \rangle}
\end{equation}
where $\langle f \rangle  = \int_\mathbb{R} t f(t) \mathrm d t$ is the mathematical expectation of the random variable with PDF $f(t)$. 
We decompose the transition density in two terms :  $T_{ji}(t,\tau) = (\mathbf 1) + ( \mathbf 2)$. The first term corresponds to the case that an edge is available to the jumper at the end of his waiting-time: 
\begin{equation}\label{eq:Tji-term1}
(\mathbf 1) = \psi_i(t-\tau) \sum_{k=1}^{|V_i|} \frac{1}{k} \binom{|V_i|-1}{k-1}p^{k}(1-p)^{|V_i|-k}. 
\end{equation}
Edge $i\rightarrow j$ has to be available, and needs to be chosen amongst the  $|V_i|-1$ other edges which are also active at  time $t$. A straightforward computation allows to rewrite equation \eqref{eq:Tji-term1}   as
\begin{equation}\label{eq:Tji-term1-bis}
(\mathbf 1) = \psi_i(t-\tau) \frac{1}{|V_i|} \left[   1-(1-p)^{|V_i|}  \right]. 
\end{equation}
The quantity between square brackets is  the probability that at least one edge is available. The factor $1/|V_i|$ appears because all outgoing edges are treated indifferently, and so the probability to be chosen is distributed uniformly amongst all edges  including $i \rightarrow j$. 

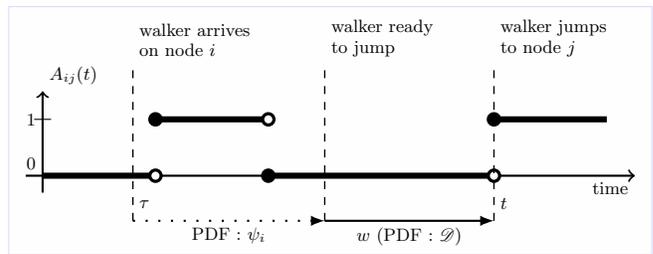
\begin{figure}[] 
	\centering
	\hspace*{-.5em}
	\begin{tikzpicture}[scale = .75,transform shape]
	%\draw[help lines,line width=1,gray,thin] (-5, -5) grid (15,15);
	\path[draw,fill = gray!1!white,draw = blue!15!white] (-.6,-.4) rectangle (10.8,4);
	% AXIS FOR A_ij 
	\begin{scope}[yshift = 1cm]
	\draw[ thick,->] (0,-.3)--(0,0)node[anchor=south east]{0} 
	--(0,1)node[anchor=east]{1}
	-- (0,1.5) node[anchor=south west]{$A_{ij}(t)$};
	\draw(-.15,1)--(.15,1);
	\draw[thick,->](-.3,0) -- (10.5,0)node[anchor = north east](t){time};
	
	% EVOLUTION OF A_ij
	%\draw[fill] (0,0)circle(.1); 
	\draw[line width = 2.3pt](0,0)--(2,0) (2,1)--(4,1) (4,0)--(8,0) (8,1)--(10,1);
	\draw[fill = white,very thick] (2,0) circle (.1);
	\draw[fill = white,very thick] (4,1) circle (.1);
	\draw[fill = white,very thick] (8,0) circle (.1);
	%\draw[fill = white,very thick] (10,1) circle (.1);
	
	\draw[fill = black,very thick] (2,1) circle (.1);
	\draw[fill = black,very thick] (4,0) circle (.1);
	\draw[fill = black,very thick] (8,1) circle (.1);
	
	\draw[dashed,line width = .5pt] (1.6,-.8) -. (1.6,2)node[anchor = south west,text width = 2cm](arrival){walker arrives on node $i$};
	\node[anchor = west] (tau) at (1.6,-.5) {$\tau$}; 
	\node[anchor = west] (t) at (8,-.5) {$t$}; 
	\draw[dashed,line width = .5pt] (5,-.8) -. (5,2)node[anchor = south west,text width = 2cm](walker){walker ready to jump};
	\draw[dashed,line width = .5pt] (8,-.8) -. (8,2)node[anchor = south west,text width = 2cm](walker){walker jumps to node~$j$};
	\draw[->,thick,>=latex] (5,-.8) --(8,-.8)node[pos = .5, anchor = north](lambda){$w$ (PDF : $\mathscr D$)};
	\draw[->,thick,>=latex,loosely dotted] (1.6,-.8) --(5,-.8)node[pos = .5, anchor = north](psi){ PDF : $\psi_i$};
	\end{scope}
	\end{tikzpicture}
	\caption{Waiting-time $w$ of the trapped walker on node $i$, before activation of edge $i \rightarrow j$. The walker arrived at time $\tau$ on node $i$. After a waiting-time determined by the density $\psi_i(t)$, a jump can be performed but none of the $|V_i|$  outgoing edges is active. The walker needs to wait a subsequent duration $w$ before a link - here to node~$j$- becomes available. So eventually the jump is performed at time $t$.   }
	\label{fig:bus-paradox}
\end{figure}

In the second case represented by figure~\ref{fig:bus-paradox}, the jump occurs after the walker happened to be trapped. Let us observe that when the walker becomes trapped on node $i$, then for a given $j \in V_i$  the time $w$ before $i \rightarrow j$ becomes available has the PDF
\begin{equation}\label{eq:PDF-theta}
\mathscr D(t) = \frac{1}{\langle D \rangle} \int_t^\infty D(\nu) \mathrm d \nu  
\end{equation}
as follows from the so-called bus-paradox.

% FIGURE : BUS PARADOX

Edge $i \rightarrow j $ is selected by the trapped walker to perform the jump a time $t$ if $(i)$ the waiting-time expires before $t$, $(ii)$ at that moment all other edges are not active and will remain inactive at least until $t$, and $(iii)$ edge $i\rightarrow j$ was also down but becomes active exactly at time $t$. It results that 
\begin{multline}
(\mathbf 2) = \int_{\tau}^{t} \psi_i(x-\tau) \big[ (1-p) \mathrm P \left \lbrace  w >t-x \right \rbrace \big]^{|V_i|-1}\\
\times (1-p) \mathscr D(t-x) \mathrm d x\\
\end{multline}
or in a slightly more compact way, 
\begin{multline}\label{eq:Tji-term2}
(\mathbf 2) = (1-p)^{|V_i|} \int_{\tau}^{t} \psi_i(x-\tau) \mathscr D(t-x)\\
\times \big[ \mathrm P \left \lbrace  w >t-x \right \rbrace \big]^{|V_i|-1}  \mathrm d x\\
\end{multline}
where
\begin{equation}\label{eq:Proba-w-bigger-than-t-x}
\mathrm P \left \lbrace  w >t-x \right \rbrace  = \int_{t-x}^\infty \mathscr D(s) \mathrm d s. 
\end{equation}
In short, we have shown that 
\begin{multline}\label{eq:Tij-compact}
T_{ji}(t,\tau) = c_1\psi_i(t-\tau) \\
+ c_2 \int_{\tau}^{t} \psi_i(x-\tau)\left [  \int_{t-x}^\infty \mathscr D(s) \mathrm d s \right]^{|V_i|-1} \mathscr D(t-x) \mathrm d x, 
\end{multline}
where $c_1$ and $c_2$ depend only on $\langle U \rangle$, $\langle D \rangle$, and  $|V_i|$. For the sake of readability, we have dropped the index $i$ due to  the node-dependence of $c_1$ and $c_2$.   Observe that the distribution of $U$ only matters through its mean, because only the mean value $\langle U \rangle$ influences the probability $p$. On the other hand, if the walker is ready to jump during a down-time, then the jumps occur directly at the end of this down-time, and so the full distribution of $D$ does matter. 

\begin{rmk} \label{rmk-acyclic}
	Having assumed an acyclic directed network allows us to consider all outgoing edges the same way. There is no possibility for the walker to backtrack to its previous step.  The time when an edge becomes  available to the walker does not depend on the arrival time of the walker on the node, and the density $\mathscr D$ of $w$ can be applied for all outgoing edges.  Indeed, if on the contrary the walker could jump across the cycle $i\rightarrow j \rightarrow i$, the probability for link $i\rightarrow j$ to still last can be large. This would induce a  bias on the next jump, giving it more chance to end up again in $j$. It results that, as stated before, $T_{ji}(t,\tau)$ depends  on the variables $t$ and $\tau$ through their difference $t-\tau$.  
\end{rmk}

\begin{rmk} \label{rmk-outdegree}
Also observe that the transition density  is the same for all $j \in V_i$. But the number of outgoing neighbors matters and appears in the transition density via the strength $|V_i|$ of  node $i$  in the underlying graph. 
\end{rmk}

\subsection{Limit cases on DAGs}\label{sec:limit-cases-DAG}
In this section we shortly discuss some particular cases listed in table~\ref{tab:limit-cases}.

\begin{table}[b]%The best place to locate the table environment is directly after its first reference in text
	\caption{\label{tab:limit-cases}%
		Selected particular cases. Here, $\delta$ means a dirac distribution in $0$ and   $ \mathcal E (x) $ stands for exponential with rate~$x$.  
	}
	\begin{ruledtabular}
	\begin{tabular}{l c c c }
	& $\psi_i(t)$ 		& $U(t)$ &     $D(t)$ \\	
	\colrule
	\rule{0pt}{3.5ex}case 1& $\delta$         &$\delta$&   $\mathcal E (\lambda) $\\  
	\rule{0pt}{3.5ex}case 2& $\mathcal E(\mu)$     &$\delta$&   $\mathcal E (\lambda) $\\ 
	\rule{0pt}{3.5ex}case 3& $\mathcal E (\mu) $         & $\mathcal E (\eta) $ &   $\mathcal E (\lambda) $ \\
\end{tabular} 
\end{ruledtabular}
\end{table}

%\begin{table}
%	\centering
%	\begin{tabular}{l|c c c }
%		& $\psi_i(t)$ 		& $U(t)$ &     $D(t)$ \\	
%		\rule{0pt}{3.5ex}case 1& $\delta$         &$\delta$&   $\mathcal E (\lambda) $\\  
%		\rule{0pt}{3.5ex}case 2& $\mathcal E(\mu)$     &$\delta$&   $\mathcal E (\lambda) $\\ 
%		\rule{0pt}{3.5ex}case 3& $\mathcal E (\mu) $         & $\mathcal E (\eta) $ &   $\mathcal E (\lambda) $ \\
%	\end{tabular} 
%	\caption{Selected particular cases. Here, $\delta$ means a dirac distribution in $0$ and   $ \mathcal E (x) $ stands for exponential with rate~$x$.  }
%	\label{tab:limit-cases}
%\end{table}

\subsubsection{Case 1}
In this case the  activation of the links is instantaneous and  so is the up-time. The down-time is exponentially distributed with rate $\lambda$ while the walker's waiting-time is again instantaneous meaning the agent is always ready to jump. It is then   straightforward to see that the waiting-time of a trapped walker before a given edge activates has density 
\begin{equation}
\mathscr D(x) = \frac{1}{1/\lambda} \int_x^\infty \lambda e^{-\lambda s} \mathrm d s = \lambda e^{-\lambda x} = D(x), 
\end{equation}
which results from the memorylessness of the exponential distribution. 
Moreover, the probability for an edge to be up at a random time is $p = 0$, and it follows that \eqref{eq:Tij-compact} is computed as
\begin{align}
T_{ji}(t,\tau) & = \left [ \mathrm P \lbrace w>t-\tau \rbrace \right ]^{|V_i|-1} \mathscr D(t-\tau) \nonumber \\
				   & = \left [ \int_{t-\tau}^\infty \mathscr D(s) \mathrm d s\right ]^{|V_i|-1} \mathscr D(t-\tau) \nonumber \\
				   & = \frac{1}{|V_i|} \lambda \,  |V_i|  \, \exp\left (-\lambda \, |V_i| \,(t-\tau)\right). 
\end{align}
The first factor results from the choice of one of the  edges (uniformly) in the underlying graph,  while the second factor shows the distribution is again exponential, with rate $\lambda \, |V_i|$. This is the density of the minimum of $|V_i|$ exponential distributions with parameter $\lambda$. Recall that $T_{ji}(t,\tau)$ depends only on the difference $t-\tau$ and on parameters of $\mathcal{ G}$. This shows  that the dynamics amounts to a Poisson CTRW on a static graph. In this sense, we recover the result of \cite{hoffmann2012generalized}. 
\begin{rmk}
	If the down-time is not exponentially distributed, it is still true that the transition density is written in terms of the density $\mathscr D_{(1),i} $ corresponding to the minimum of $|V_i|$ independent random variables with density  $\mathscr D$: 
	\begin{equation}\label{eq:Tji-minimum}
  T_{ji}(t,\tau) = \frac{1}{|V_i|} \mathscr D_{(1),i}(t-\tau). 
 \end{equation} 
	It is a  straightforward calculation to see that 
	\begin{align}\label{eq:Dmin}
\mathscr D_{(1),i} (t) &= -\frac{\mathrm d}{\mathrm dt} \left[ \mathrm P (w > t) \right]^{|V_i|}  \nonumber \\
&=  |V_i| \left(1-F_w(t)  \right)^{|V_i|-1}\mathscr D (t) ,
\end{align}
	where $F_w (t) $ is the distribution function of the variable $w$ with density $\mathscr D$. 
\end{rmk}

\subsubsection{Cases 2 and 3}
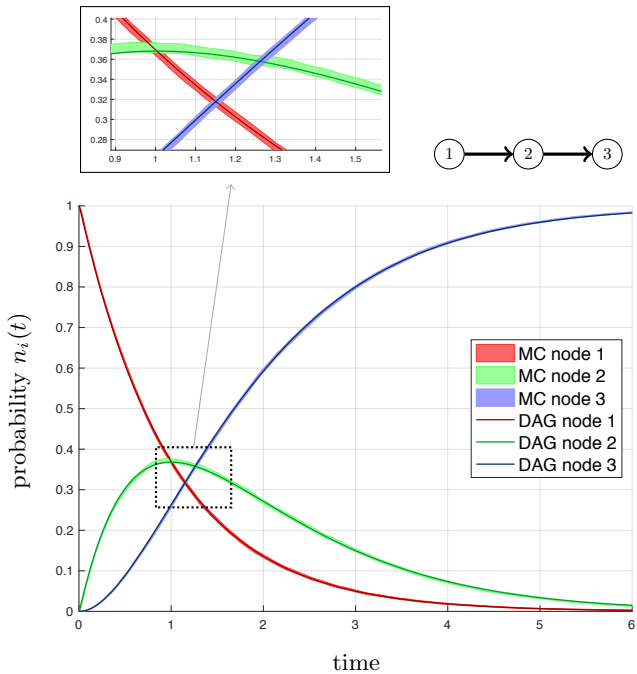
\begin{figure}[]
	\centering
\hspace*{-1em}\begin{tikzpicture}
\node (fig) {\includegraphics[width = .9\columnwidth]{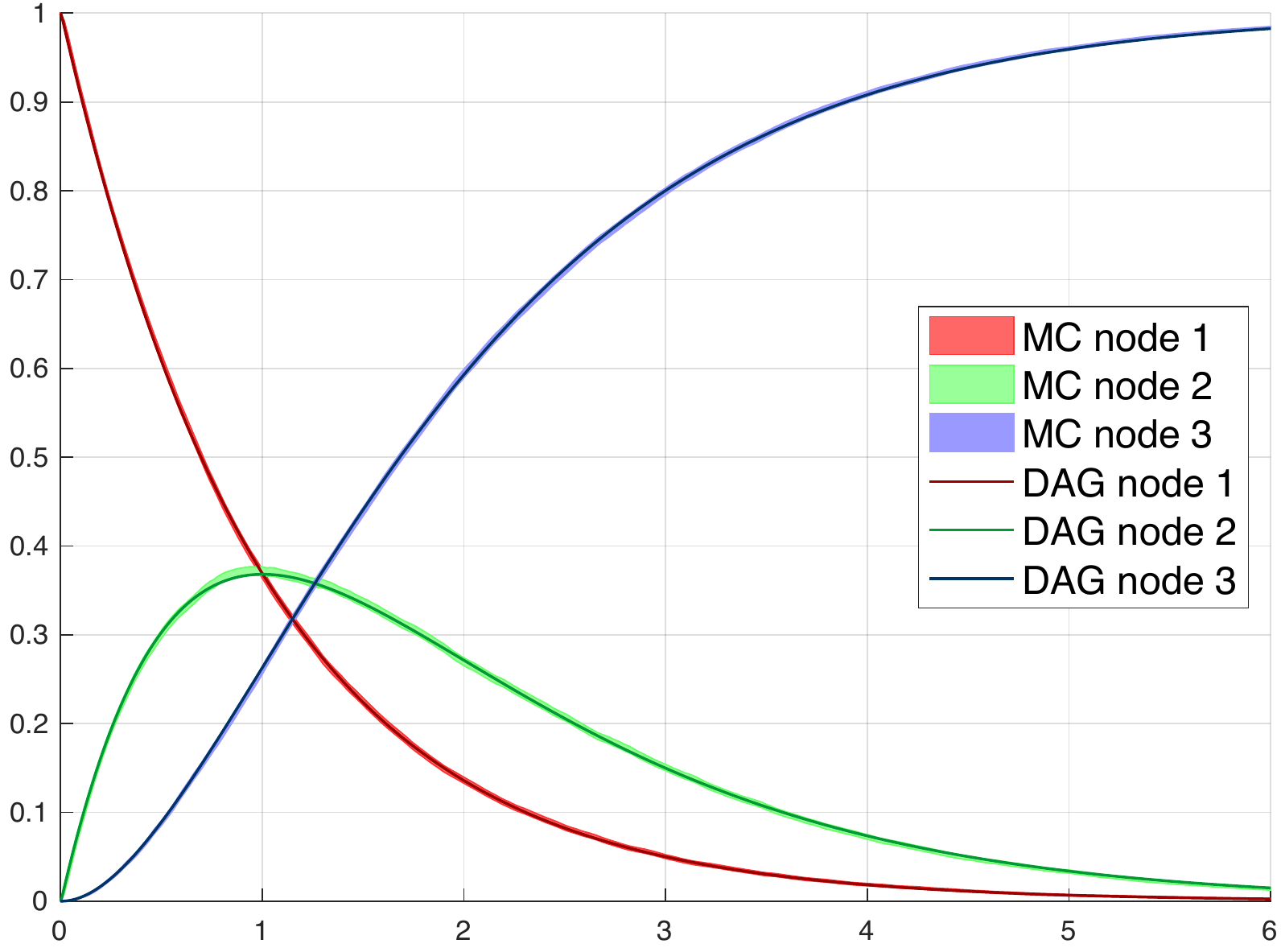}}; 
\node[below = .1em of fig,xshift = .5em](xlablel){time}; 
\node[above left = -1.5  and .1em of fig, rotate  = 90](ylablel){probability $n_i(t)$}; 
\node[draw, anchor = south west](inset)at (-3.51,3.25){\includegraphics[width= 0.45\columnwidth]{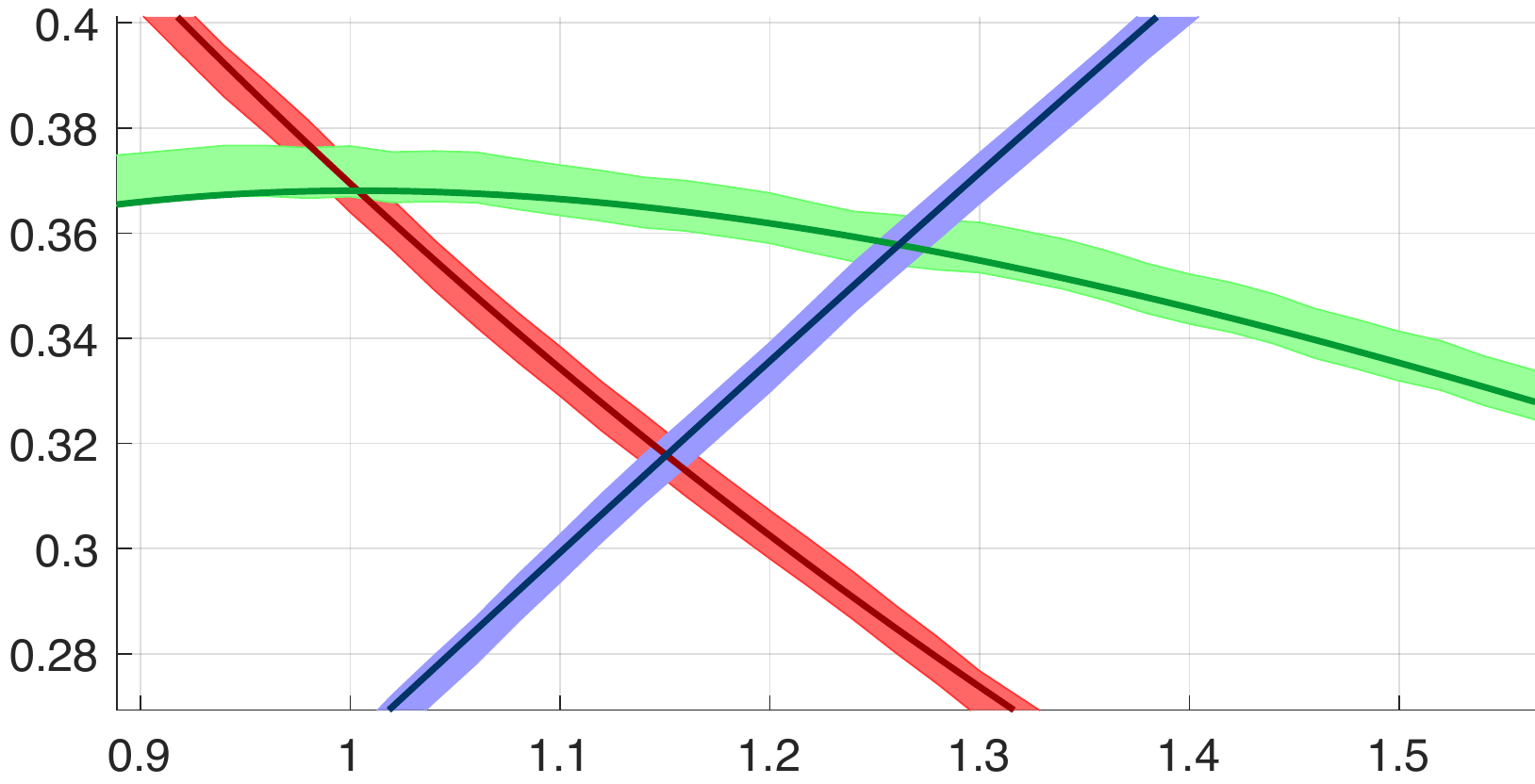}};

\begin{scope}[scale  = .7, transform shape,yshift = 5cm,xshift = 2cm]
\node[style = circle, draw] (n1) at (0,0) {1};
\node[style = circle, draw] (n2) at (1.5,0) {2};
\node[style = circle, draw] (n3) at (3,0) {3};
\draw[very thick,->] (n1) to  (n2); 
\draw[very thick,->] (n2) to  (n3);
\end{scope}

\draw[densely dotted, thick] (-2.5,-1.2) rectangle (-1.5,-0.4);
\draw[->,gray!65!white](-2,-0.4) to (-1.5,3.1); 

%\draw[step=.5cm,red!14!white] (-4.4,-4.4) grid (4.4,4.4);
%\foreach \x in {-4,-3.5,...,4}
%\draw[red] (\x cm,1pt) -- (\x cm,-1pt) node[anchor=north] {{\tiny $\x$}};
%\foreach \y in {-4,-3.5,...,4}
%\draw[red] (1pt,\y cm) -- (-1pt,\y cm) node[anchor=east] {{\tiny $\y$}};
\end{tikzpicture}

\caption{Validation of the analytical model on a DAG in case~3 of table~\ref{tab:limit-cases}. All densities are exponential : $\psi \sim \mathcal{E}(\mu = 1)$, $U \sim \mathcal{E}(\eta = 1)$, ${D \sim \mathcal{E}(\lambda = 1)}$. The Monte-Carlo simulation is the average of $10^5$  independent trajectories of a single walker. The continuous curves and markers represent the probabilities $n_i(t)$ while the errorbars stand for the standard deviation of the mean. The inset magnifies the central part of the main figure. The result of the analytical curves  are plotted with filled markers. The graph is a directed chain with three nodes, as shown in the top right corner.}
\label{fig:Case3Acyclic}
\end{figure}

In these two cases, the computation of $T_{ji}$ and $\Phi_i$ yield compact expressions (see appendix~\ref{app:case2} and~\ref{app:case3}). 
The exactness of the expressions result from the network being acyclic. An integration of the analytical model  is compared against Monte-Carlo simulation on figure~\ref{fig:Case3Acyclic} in case 3 (all exponential densities).

\subsection{Equivalent node- and edge-centric models}\label{sec:modelreduction}
In all possible cases (thus beyond exponential distributions), the model for DAGs  can be cast into a nodes-only process on a static network, or to an edges-only process with instantaneous edges activation, and a walker with no waiting-time. 

In the former case for instance, only a waiting-time density of the walker is retained, and it can be computed from the densities $\psi$, $U$ and $D$ of the original model. For the sake of compactness, we assume all edges to follow the same densities. The all-in-one waiting-time PDF for the walker in node $i$ with $|V_i|>0$ is 
\begin{equation}\label{eq:reduced-Psi}
\Psi_i(t) = (\psi_i * \widetilde{\mathcal{D}} _{(1),i})(t)
\end{equation} 
where $*$ denotes a convolution in the time variable and with 
\begin{equation}\label{eq:reduced-D}
\widetilde{\mathcal{D}} _{(1),i}(t) =  (1-p)^{|V_i|} \mathcal{D}_{(1),i}(t) + (1-(1-p)^{|V_i|}) \delta(t). 
\end{equation}
It results that 
\begin{equation}\label{eq:reduced-Psi2}
\Psi_i(t) = (1-p)^{|V_i|} (\psi_i *\mathcal{D}_{(1),i})(t) + (1-(1-p)^{|V_i|}) \psi_i(t).
\end{equation}

The model reduction in the edge-centric case can be deduced from this formula. Let $X_i$ be the random variable with density $\Psi_i$ and let $Y_{i\bullet}$ be the random variable for the   waiting-time associated in the reduced model to an edge originating from node $i$ with degree $|V_i|>0$. Then $X_i$ is the minimum of $|V_i|$ i.i.d. random variables such as $Y_{i\bullet}$ and we know 
\begin{equation}\label{eq:reduced-FX}
F_{X_i}(t) = 1-(1-F_{Y_{i\cdot}})^{|V_i|}. 
\end{equation}
yielding
\begin{equation}\label{eq:reduced-FY}
F_{Y_{i\bullet}}(t) = 1-(1-F_{X_i}(t))^{1/|V_i|}
\end{equation}
The PDF $\Delta_{i\bullet}$ of the waiting-time on the edge can be obtained via 
\begin{equation}\label{eq:reduced-Delta}
\Delta_{i\bullet}(t) = \frac{1}{|V_i|} (1-F_{X_i}(t))^{1/|V_i|} \Psi_i(t). 
\end{equation}

\section{Cycles and emergence of  memory \label{sec:emergenceMemory}}
\begin{figure}
	\centering
	\hspace*{-.5em}\begin{tikzpicture}
	\node (fig1) {\includegraphics[height = .38\columnwidth]{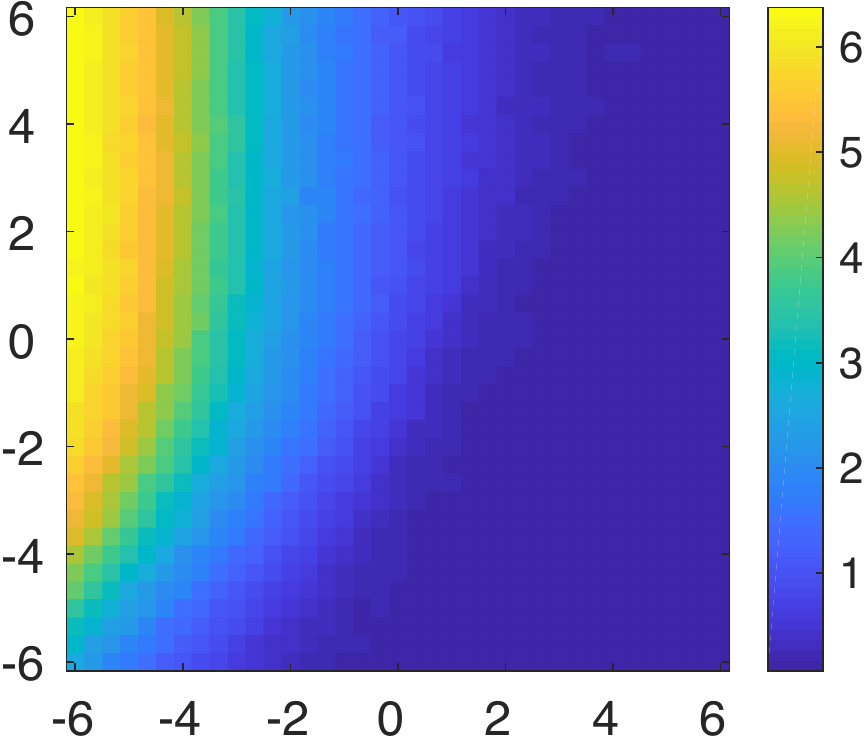}};  %
	\node[right = 0.3cm of fig1] (fig2) {\includegraphics[height = .38\columnwidth]{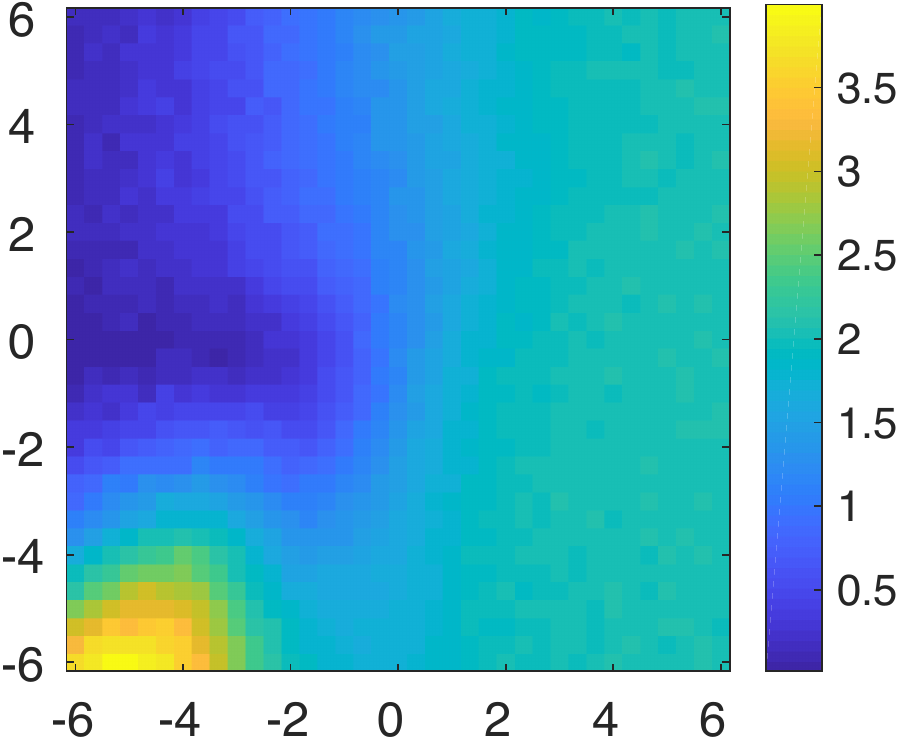}}; 
	
	\node[below = .1em of fig1](xlablel){down-time rate $(\log_2 \lambda)$}; 
	\node[below = .1em of fig2](xlablel){down-time rate $(\log_2 \lambda)$}; 
	
	\node[above left = -0.1  and -.3em of fig1, rotate  = 90](ylablel){up-time rate $(\log_2 \eta)$};

	\begin{scope}[scale  = .8, transform shape, xshift = 1.2cm, yshift = -4.3cm]
	\node[style = circle, draw] (n1) at (0,0) {1};
	\node[style = circle, draw] (n2) at (1.5,0) {2};
	\node[style = circle, draw] (n3) at (3,0) {3};
	\draw[very thick,->] (n1) to  (n2); 
	\draw[very thick,->] (n2) to[bend left]  (n1); 
	\draw[very thick,->] (n2) to  (n3);
	\draw[very thick,->] (n3) to[bend right]  (n1);
	\end{scope}
	
	\begin{scope}[scale = .5]
	\node[style = circle, draw,fill = white,inner sep = 2pt] (A) at (-2.3,-1.8) {{\scriptsize A}}; 
	\node[style = circle, draw,fill = white,inner sep = 2pt] (D) at (-2.3,2.3) {{\scriptsize D}}; 
	\node[style = circle, draw,fill = white,inner sep = 2pt] (B) at (1.7,-1.8) {{\scriptsize B}};
	\node[style = circle, draw,fill = white,inner sep = 2pt] (C) at (1.7,2.3) {{\scriptsize C}};
	\end{scope}
	
	\begin{scope}[scale = .5,xshift = 1.02\columnwidth]
	\node[style = circle, draw,fill = white,inner sep = 2pt] (A) at (-2.3,-1.8) {{\scriptsize A}}; 
\node[style = circle, draw,fill = white,inner sep = 2pt] (D) at (-2.3,2.3) {{\scriptsize D}}; 
\node[style = circle, draw,fill = white,inner sep = 2pt] (B) at (1.7,-1.8) {{\scriptsize B}};
\node[style = circle, draw,fill = white,inner sep = 2pt] (C) at (1.7,2.3) {{\scriptsize C}};
	\end{scope}
	
	\node[above = .2cm of fig1,text width = .4\columnwidth,align = center]{(a)\\active  node-centric};
	
		\node[above = .2cm of fig2,text width = .4\columnwidth,align = center]{(b)\\passive  edge-centric};

	\end{tikzpicture}
	
	\caption{Comparison of the classical active node-centric (left panel) and passive edge-centric (right panel) models with a Monte-Carlo simulation involving $5\cdot 10^3$ independent trajectories. The errors $E_\mathrm{active}(T)$ and $E_\mathrm{passive}(T)$   between  the predictions  of the models and the actual (Monte-Carlo) probabilities $n_i(t)$ as given by equation~\eqref{eq:error-models}, is plotted for various combinations  of the rates of the exponential up-time and down-time densities $\eta$ and $\lambda$. The walker's exponential density has rate $\mu = 1$ on all nodes. Each of the four  regimes marked by the letters $A$ to $D$ correspond to the four scenarios previously identified on figure~\ref{fig:cartoon}. Section~\ref{sec:emergenceMemory} aims at providing the necessary modeling framework to cover the full domain of this plot, and to even go beyond the case considered here,   allowing not clearly separated timescales.   The graph appears at the bottom of the figure, and the duration of the simulations is $T=10$.   }
	\label{fig:numericalError}
\end{figure}
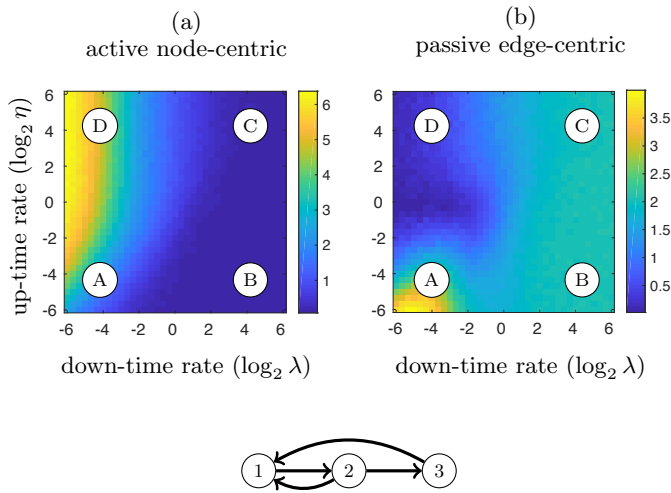

The random walk under scrutiny in this work involves three processes, each with its own timescale and characterized by the densities $\psi_i$, $U$ and $D$. Section~\ref{sec:model} and figure~\ref{fig:cartoon} in particular offered a qualitative evidence of three possible scenarios. 
In the first one,  the durations of the down-times are fast with respect to the typical walkers' waiting-time, and node-centric modeling proves applicable. In the second one, the down-times (resp. the up-times) are relatively slow (resp. fast) as compared to the walker, and edge-centric modeling is effective. In the third scenario however, when none of the two previous assumptions holds true, the modeling needs not neglect any of three processes. This claim is hereby sustained  by figure~\ref{fig:numericalError} where the evolution of $\mathbf{n}_{\mathrm{Monte-Carlo}}(t)$ from $5\cdot 10^3$ Monte-Carlo simulations  is compared with the predictions from the active node-centric and  the passive edge-centric models, in the all-exponential case. In the former model, the dynamics of the edges is neglected~: a static network is assumed and the master equation is
\begin{equation}\label{eq:master-node-centric}
\dot{\mathbf{n}}_{\mathrm{active}} = - \mathbf{n}_{\mathrm{active}} \Big(      I-\mathrm{diag}\big(|V_1|, \ldots, |V_N| \big)^{-1} G      \Big)
\end{equation}
where $G$ is the adjacency matrix of the underlying network $\mathcal{ G}$ and the time is scaled by the rate $\mu$ of the walker. 
In the latter case, the walker has no own waiting-time. The inter-activation dynamics of the edges is accounted for, while the activations are  instantaneous. Therefore, the time is scaled according to the rate $\lambda$ of the down-times and
\begin{equation}\label{eq:master-edge-centric}
\dot{\mathbf{n}}_{\mathrm{passive}} = - \mathbf{n}_{\mathrm{passive}} \Big(    \mathrm{diag}\big(|V_1|, \ldots, |V_N|\big)  - G      \Big). 
\end{equation}
The norm of the error between the numerical simulation and the two models is then integrated over the duration $T$  of the simulations, 
\begin{equation}\label{eq:error-models}
E_\mathrm{model}(T) = \int_0^{T}   \big \Vert       \mathbf{ n}_\mathrm{model}(s)  - \mathbf{ n}_\mathrm{Monte-Carlo}(s)           \big \Vert_2 \mathrm d s,
\end{equation}
where ``model" stands for ``active" or ``passive". Note that in the three preceding equations, $\mathbf{ n}_{\mathrm{model}}$ is a row vector.  The outcome is represented by  figure~\ref{fig:numericalError}
in the $(\log_2 \lambda ,  \log_2 \eta)$ plane, having chosen a  rate $\mu = 1$ for $\psi_i$ on the nodes. 

\begin{figure}
\centering
	\hspace*{-1em}%\includegraphics[width = .85\columnwidth]{memoeffect2.pdf}
	\begin{tikzpicture}
	\node (fig) {\includegraphics[width = .8\columnwidth]{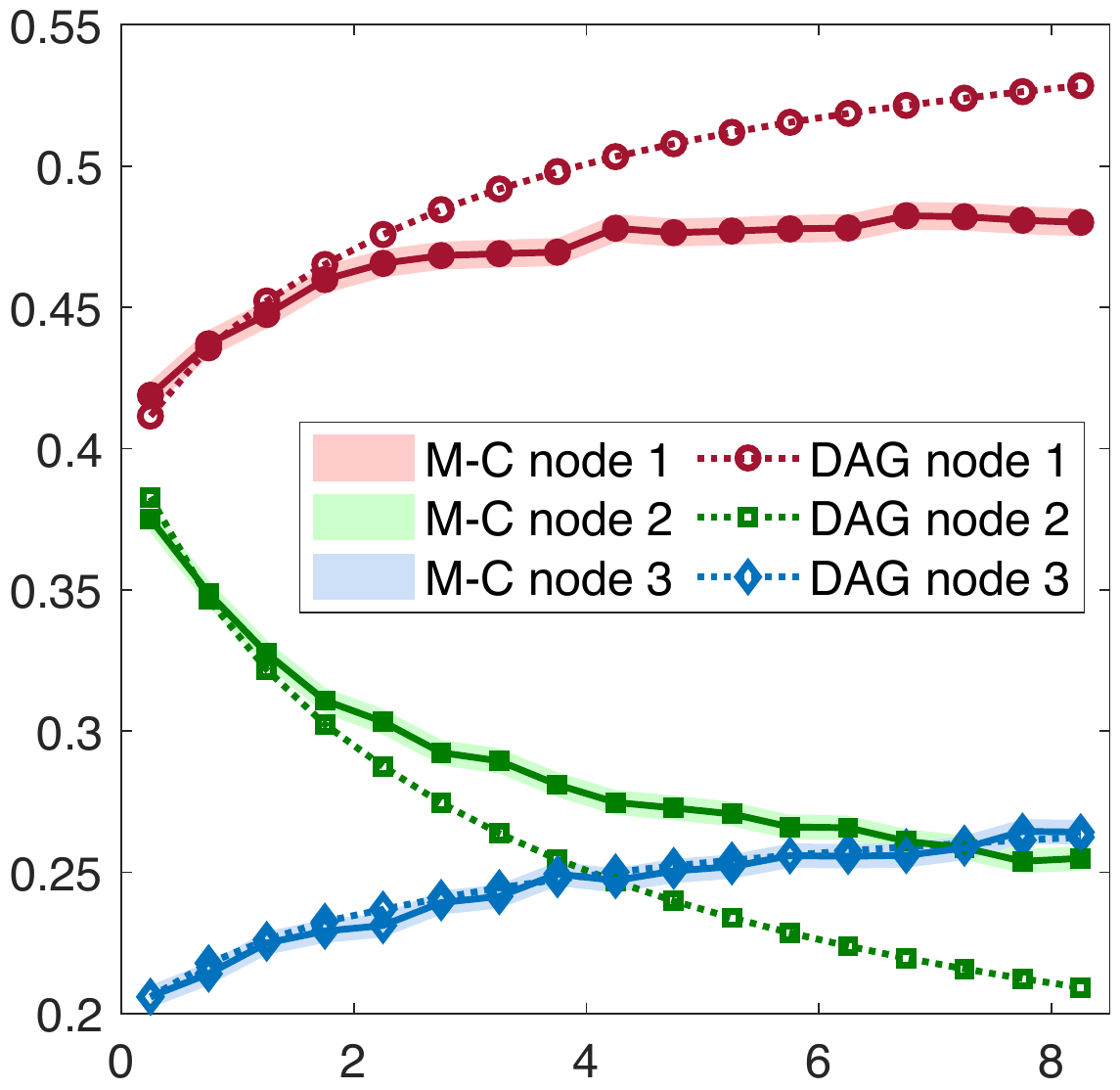}}; 
	\node[below = -.1em of fig](xlablel){$\quad$ walker rate $\mu$ }; 
	\node[above left = -1.2  and -.1em of fig, rotate  = 90](ylablel){asymptotic state $n_i(t\rightarrow \infty)$};
	
	% add line on monte carlo shading
	\draw[red,thick] (-1.52,.53) --(-0.91,0.53);
	\draw[green,thick] (-1.52,.16) --(-0.91,0.16);
   \draw[blue,thick] (-1.52,-0.19) --(-0.91,-0.19);
	
%	\draw[step=.5cm,red!14!white] (-4.4,-4.4) grid (4.4,4.4);
%	\foreach \x in {-4,-3.5,...,4}
%	\draw (\x cm,1pt) -- (\x cm,-1pt) node[anchor=north] {{\tiny $\x$}};
%	\foreach \y in {-4,-3.5,...,4}
%	\draw (1pt,\y cm) -- (-1pt,\y cm) node[anchor=east] {{\tiny $\y$}};
	  
	\end{tikzpicture}
	\caption{The formula's for DAGs   are no longer valid if there are cycles, as can be seen from this comparison with Monte-Carlo simulations. On this figure, the stationary state $n_i(t\rightarrow \infty)$ in each node for varying  values of the rate $\mu$ of the exponential density of the walker's waiting-time $\psi(t)$ is plotted resulting from Monte-Carlo simulation (solid lines with filled markers) and the analytical model on DAGs (dotted lines with empty markers). The width of the shading around the Monte-Carlo curves corresponds to   twice the standard deviation of the mean computed on $4\cdot 10^4$ independent trajectories.  The up- and down-time also follow exponential densities,  $U \sim \mathcal{E}(\eta = 1),  D \sim \mathcal{E}(\lambda = 1)$, and the initial condition of the walk is $\mathbf{ n}(0) = (1,0,0)^T$.  This example illustrates that when  the timescale of the walker is faster, the memory effect becomes more pronounced and the error with respect to the Monte-Carlo simulations increases. The graph is the one of figure~\ref{fig:numericalError}.}
	\label{fig:cycle_vs_DAG}
\end{figure}
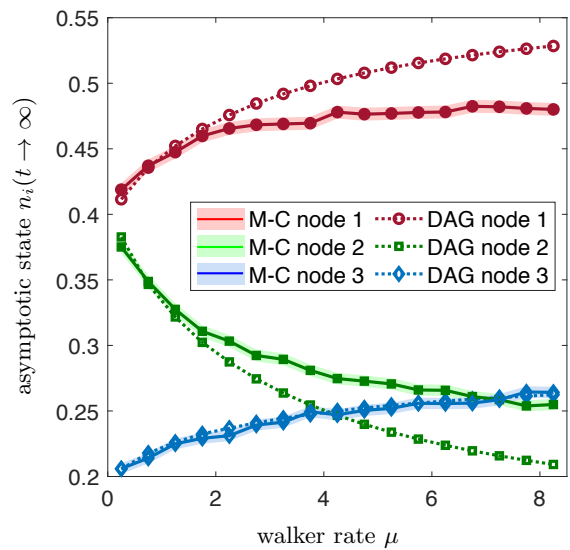

The region where  both errors are large demonstrates the need for the inclusive model developed  in section~\ref{sec:reducible-model}, where the full interplay of the walker's and edges behaviors are accounted for. The results derived thus far relied  on an assumption of independence between events, \emph{i.e.} links creation and destruction, encountered by the random walker. 
This assumption is  clearly valid for DAGs but ceases to hold true when  the underlying network has cycles. In that case, the walker may be influenced by the statistical information left at the previous passage, which may induced biases in the walker trajectory \cite{Speidel2015PhysRevE,gueuning2017backtracking}. 
The acyclic predictions are however expected to remain good approximations  if the process on the nodes ($\psi$) is slow with respect to the edges dynamics, either in the case of long cycles or also locally  if  nodes have high degree. 
In other cases, as illustrated in figure~\ref{fig:cycle_vs_DAG}, one can observe significant deviations between the approximation and the numerical simulations of the process, even in situations when each of the three processes is a Poisson process. In such cases, we will observe the emergence of memory, or loss of the Markov property, in the trajectories of the walker.

In general, if cycles are present in the network, the state space is the full trajectory of the random walk, which makes the problem intractable analytically. We hereby propose  a method estimating the corrections due to cycles of a given length, and which generalizes the results in section \ref{sec:reducible-model}. Although the proposed framework is general, we restrict the following discussion to contributions of cycles of length 2. This choice is motivated by the sake of simplicity and speeds up  numerical simulations, as the incorporation of long cycles  comes with increased computational cost.  Also note that  longer  cycles are associated to  weaker  corrections, as more time between two passages tends to wash out  footprints left by the walker.

\subsection{Master equation with corrections for $\mbox{2-cyles}$}
\label{sec:masterequation_cycles}
We need to enlarge the state space of the system in order to allow a correction for $\mbox{2-cycles}$. Let us accordingly first define $q_{imm'}(\tau,\nu)$ to be the arrival time  density for the couple $(\tau,\nu)$ on nodes $m' \rightarrow m \rightarrow i$. Observe that almost surely, $0  < \nu < \tau$. As depicted by figure \ref{fig:jumptimes1}, let  $T_{j|imm'}(t|\tau,\nu)$ be the conditional transition density across edge $i\rightarrow j$ at time $t$, taking the two previous jumps into account : from $m'$ to $m$ at time $\nu$ and from $m$ to $i$ at time $\tau$. It will become clear that by the limited amount of memory we take into account, this conditional density actually only depends on the durations $t-\nu$ and $ \tau-\nu$.     Let also $\Phi_{imm'}(t|\tau,\nu)$ be the probability to stay up to time $t$ on node $i$, having arrived at time $\tau$ in the node, and having made the two previous jumps at times $\nu \leq \tau$  as represented by  figure~\ref{fig:stayput1}. 
\begin{figure}
	\centering
	\begin{tikzpicture}[scale = 1,transform shape]
		\path[draw,fill = white] 
		(0,0) node[style = circle,draw,text width = 1.3em,align = center](c1){$m'$}
		++(1.5,0)   node[style = circle,draw,text width = 1.3em,align = center](c2){$m$}
		++(1.5,0)  node[style = circle,draw,text width = 1.3em,align = center](c3){$i$}
		++(1.5,0)  node[style = circle,draw,text width = 1.3em,align = center](c4){$j$};
		\coordinate (c0) at (-1,.5); 
		
		% add arrows with jump times
		%\draw[line width = 1pt,->] (c0) to[bend left] node[pos=.5,above]{$\omega$} (c1);
		\draw[line width = 1pt,->] (c1) to[bend left] node[pos=.5,above]{$\nu$} (c2);
		\draw[line width = 1pt,->] (c2) to[bend left] node[pos=.5,above]{$\tau$} (c3);
		\draw[line width = 1pt,->] (c3) to[bend left]node[pos=.5,above]{$t$}  (c4);
	\end{tikzpicture}
	\caption{Jump times and nodes in the definition of the transition density $T_{j|imm'}(t|\tau, \nu)$. The arrows are labeled by the jump time. Here, nodes $m'$ and $i$ and nodes $m$ and $j$ are not necessarily different nodes.  }
	\label{fig:jumptimes1}
\end{figure}
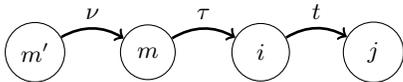
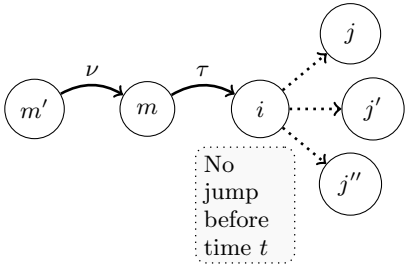
\begin{figure}
	\centering
	\begin{tikzpicture}[scale = 1,transform shape]
	\path[draw,fill = white] 
	(0,0)  node[style = circle,draw,text width = 1.3em,align = center](c1){$m'$}
	++(1.5,0)  node[style =circle,draw,text width = 1.3em,align = center](c2){$m$}
	++(1.5,0)  node[style = circle,draw,text width = 1.3em,align = center](c3){$i$}
	+(1.2,1) node[style = circle,draw,text width = 1.3em,align = center](c4){$j$}
	+(1.5,0) node[style = circle,draw,text width = 1.3em,align = center](c5){$j'$}
	+(1.2,-1) node[style = circle,draw,text width = 1.3em,align = center](c6){$j''$};
	\coordinate (c0) at (-1,.5); 
	
	% add arrows with jump times
	%\draw[line width = 1pt,->] (c0) to[bend left] node[pos=.5,above]{$\omega$} (c1);
	\draw[line width = 1pt,->] (c1) to[bend left] node[pos=.5,above]{$\nu$} (c2);
	\draw[line width = 1pt,->] (c2) to[bend left] node[pos=.5,above]{$\tau$} (c3);
	\draw[line width = 1pt,->,dotted] (c3) to[] (c4);
	\draw[line width = 1pt,->,dotted] (c3) to[] (c5);
	\draw[line width = 1pt,->,dotted] (c3) to[] (c6);
	\path (2.8,-.5)node[anchor = north, text width = 1.1cm,,fill = gray!5!white,rounded corners,draw,dotted,line width = .5pt] (txt){No jump before time $t$};
	\end{tikzpicture}
	\caption{Jump times and nodes in the definition of the probability $\Phi_{imm'}(t|\tau, \nu)$. The arrows are labeled by the jump time. Nodes $m'$ and $i$ could be the same node. }
	\label{fig:stayput1}
\end{figure}

We have 
\begin{equation}\label{eq:stayput2}
	\Phi_{imm'}(t|\tau, \nu) = 1 - \sum_{j \in V_i} \int_\tau^t T_{j|imm'}(s|\tau,\nu)\, \mathrm ds. 
\end{equation}
The normalization condition reads
\begin{equation}\label{eq:normalizationPhiimmprime}
\lim_{t\rightarrow \infty }\Phi_{imm'}(t|\tau,\nu) = 0, \quad \forall \ 0\leq \nu \leq \tau, 
\end{equation}
and so
\begin{equation}\label{eq:normalizationTjimmprime}
\sum_{j \in V_i} \int_\tau^\infty T_{j|imm'}(s|\tau,\nu)\, \mathrm ds = 1
\end{equation}
for all $0 \leq \nu \leq \tau$ and $1\leq i \leq N$. In the remainder of this section the computations assume the conditional transition density  to be known. Its exact form will be determined in the next section. 

Using the same steps as for acyclic graphs, let us first write the probability that the walker is on node~$i$ at time~$t$ as 
\begin{equation}\label{eq:n-decomposed}
n_i(t) = n_i^{(0)}(t) + n_i^{(1)}(t) + n_i^{(k \geq 2)}(t)
\end{equation} 
where the superscript refers to  the number of jumps performed up to time $t$. The first two terms  are not impacted by the memory effect, and can be computed based on the transition densities established under the no-cycle hypothesis : 
\begin{align}
n_i^{(0)}(t) &= \int_0^t q_i^{(0)}(t) \Phi_i(t,\tau) \mathrm d \tau \nonumber \\
                 & = n_i(0) \Phi_i(t,0), 
\end{align}
and 
\begin{align}
n_i^{(1)}(t) &= \int_0^t q_i^{(1)}(\tau) \Phi_i(t,\tau) \mathrm d \tau  \nonumber \\
& = \sum_{m \in V'_i} n_m(0) \int_0^t T_{im}(\tau,0) \Phi_i(t,\tau) \mathrm d \tau . 
\end{align}
It remains to compute $n_i^{(k\geq 2)}(t) = \sum_{k \geq 2} n_i^{(k)}(t)$. 
Note that in $n_i^{(k)}(t)$ we also need the transition density of the \mbox{$(k+1)$-th} jump which determines the probability to stay put on node $i$ up to time $t$ after $k$ jumps. For all $k \geq 2$ one can write
\begin{multline}
n_i^{(k)} (t) = \sum \sum_{\hspace*{-2em}m' \rightarrow m \rightarrow i} \ \  \iint \limits_{0  \leq \nu \leq \tau } q_{imm'}^{(k,k-1)} (\tau, \nu ) \\ 
\times \Phi_{imm'}(t| \tau,\nu) \,  \mathrm d \nu \mathrm d \tau, 
\end{multline}
where again the superscript in $q_{imm'}^{(k,k-1)} $ gives the number of jumps. In order to determine $n_i^{(k\geq 2)} (t)$ we will  need 
\begin{equation}\label{eq:q_tuple}
q_{imm'} (\tau, \nu) = \sum_{k \geq 2} q_{imm'}^{(k,k-1)} (\tau, \nu). 
\end{equation}
Once we have computed this quantity, then the third term in \eqref{eq:n-decomposed},  $n_i(t) = n_i^{(0)}(t) + n_i^{(1)}(t) + n_i^{(k \geq 2)}(t)$, will indeed follow as
\begin{multline}\label{eq:ni-with-cycles}
n_i^{(k \geq 2)} (t) = \sum \sum_{\hspace*{-2em}m' \rightarrow m \rightarrow i} \ \  \iint \limits_{0  \leq \nu \leq \tau } q_{imm'} (\tau, \nu ) \\ 
\times \Phi_{imm'}(t| \tau,\nu) \,  \mathrm d \nu \mathrm d \tau. 
\end{multline}
and we have obtained the probability $n_i(t)$ in function of the initial condition $\mathbf{n}(0)$. 

Let us therefore  determine the arrival-times density in a given number of jumps, $q_{imm'}^{(k,k-1)}(\cdot,\cdot)$.  Let us write equation \eqref{eq:q_tuple} by splitting the sum as 
\begin{multline}\label{eq:q_tuple2}
q_{imm'}(\tau,\nu) =  \sum_{k = 2}^{\infty} q_{imm'}^{(k+1,k)}(\tau,\nu) 
+ q_{imm'}^{(2,1)}(\tau,\nu). 
\end{multline}
In this expression, for all $k \geq 2$, 
\begin{multline}
q_{imm'}^{(k+1,k)}(\tau,\nu) = \sum_{m''\in V'_{m'}} \int_0^{\nu}T_{i|mm'm''}(\tau|\nu,\nu')\\ 
\times q_{mm'm''}^{(k,k-1)}(\nu,\nu')\,  \mathrm d \nu'
\end{multline}
and using again \eqref{eq:q_tuple}, equation \eqref{eq:q_tuple2} becomes 
\begin{multline}\label{eq:q_tuple3}
q_{imm'}(\tau,\nu) =  \sum_{m'' \in  V'_{m'}} \int_0^{\nu}T_{i|mm'm''}(\tau|\nu,\nu')\\
\times q_{mm'm''}(\nu,\nu')\, \mathrm d \nu' + q_{imm'}^{(2,1)}(\tau,\nu). 
\end{multline}
The extended  initial condition of arrival times for the first two jumps is given by 
\begin{align}\label{eq:extended-init-cond}
q_{imm'}^{(2,1)}(\tau,\nu) &= \widetilde{T}_{im} (\tau-\nu)  \int_0^{\nu} \widetilde{T}_{mm'}(\nu-\nu')q_{m'}^{(0)}(\nu') \mathrm d \nu' \nonumber\\
											&=\widetilde{T}_{im} (\tau-\nu) \widetilde{T}_{mm'}(\nu) n_{m'}(0)
\end{align}
where $\widetilde{T}_{ji}(t) := T_{ji}(t,0)$ is the transition density for the acyclic case.

Equation \eqref{eq:q_tuple3} is a Volterra linear integral equation of the second kind, with kernel given by the conditional transition density that is determined hereafter.  We have a vector of unknown functions $\mathbf Q $, where each component function $q_{imm'} (\cdot,\cdot): [0,\infty)^2\rightarrow [0,\infty)  $ corresponds to a  path of length 2 in the underlying graph $\mathcal G$. As will appear clearly in the sequel, this equation cannot be cast under the form of a convolution, because as we will see  $T_{i|mm'm''}(\tau|\nu,\nu') = T_{i|mm'm''}(\tau-\nu'|\nu-\nu',0)$.  Consequently, the Laplace-transform-based method cannot be applied.

\subsection{Transition density with correction for $\mbox{2-cycles}$}
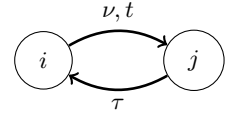
\begin{figure}
	\centering
	\begin{tikzpicture}[scale = 1,transform shape]
	\path[draw,fill = white] 
	(0,0) node[circle,draw,text width = 1.3em,align = center](c1){$i$}
	++(2,0) node[circle,draw,text width = 1.3em,align = center](c2){$j$}; 
	\coordinate (c0) at (-1,.5); 
	
	% add arrows with jump times
	%\draw[line width = 1pt,->] (c0) to[bend left] node[pos=.5,above]{$0$} (c1);
	\draw[line width = 1pt,->] (c1) to[bend left] node[pos=.5,above]{$\nu,t$} (c2);
	\draw[line width = 1pt,->] (c2) to[bend left] node[pos=.5,below]{$\tau$} (c1);
	\end{tikzpicture}
	\caption{Jump times and edges through which the jumps occur  in the transition density ${T}_{j|iji}(t|\tau,\nu)$. The arrows are labeled by the jump time. Note that node $i$ can possibly have other out-neighbors than $j$. The corresponding edges would then impact the transition density through edge $i\rightarrow j$.}
	\label{fig:cycle_case2}
\end{figure}

\begin{figure}
	\centering
	\begin{tikzpicture}[scale = 1,transform shape]
	\path[draw,fill = white] 
	(0,0)  node[circle,draw,text width = 1.3em,align = center](ci){$i$}
	+(1.8,0)  node[circle,draw,text width = 1.3em,align = center](cj){$j$}
	+(-1.8,0)    node[circle,draw,text width = 1.3em,align = center](cm){$m$}; 
	\coordinate (c0) at (-1,.5); 
	
	% add arrows with jump times
	%\draw[line width = 1pt,->] (c0) to[bend left] node[pos=.5,above]{$0$} (c1);
	\draw[line width = 1pt,->] (ci) to[bend right] node[pos=.5,above]{$\nu$} (cm);
	\draw[line width = 1pt,->] (cm) to[bend right] node[pos=.5,below]{$\tau$} (ci);
	\draw[line width = 1pt,->] (ci) to[bend left] node[pos=.5,above]{$t$} (cj);
	\end{tikzpicture}
	\caption{Jump times and edges corresponding to the jumps  in the transition density $ {T}_{j|imi}(t|\tau,\nu)$. The arrows are labeled by the jump time. Here, $m$ and $j$ are assumed to be different nodes. Not all  out-neighbors of node $i$ are represented, although they would influence the transition density. }
	\label{fig:cycle_case3}
\end{figure}
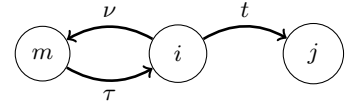

We want to compute $T_{j|imm'}(t|\tau,\nu)$.  The trajectory before the jump at time $\nu$ is not taken into account and so only durations starting from time $\nu$ matter : 
\begin{equation}
T_{j|imm'}(t|\tau,\nu) = T_{j|imm'}(t-\nu|\tau-\nu,0). 
\end{equation}
Therefore, we need to determine $\widetilde{T}_{j|imm'}(x|y):=T_{j|imm'}(x|y,0)$, $0 \leq y \leq x $. 
There are three cases, depending on whether $(m' \rightarrow m \rightarrow i )$ is a 2-cycle or not.  
\begin{itemize}
\item In the first case, $m' \neq i$, and there is no memory effect due to 2-cycles. The density reads as before
\begin{equation}\label{eq:factorization}
\widetilde{T}_{j|imm'}(x|y) = \widetilde{T}_{ji}(x-y) \, 
\end{equation}
where the right-hand side is the one   from the modeling for DAGs. 
\item In the second case, $(m',m) = (i,j)$ and we have the situation depicted by figure \ref{fig:cycle_case2}. The density cannot be written in terms of the one  obtained for acyclic graphs. 
\item In the third case, $m' = i$ but $m\neq j$, as shown in figure~\ref{fig:cycle_case3}, and again, we do not have a reduction like in \eqref{eq:factorization}. 
\end{itemize}
By definition, $ \widetilde{T}_{j|imm'}(x|y)  =  T_{j|imm'}(x|y,0)$ with $x = t-\nu$ and $y = \tau-\nu$. In the following, the letters $t,\tau, \nu$ will indicate absolute times, whereas $x$ and $y$ are durations. We will keep both in order to avoid having to assume a jump a time $0$. As before, in the second and in the third case, we will write
\begin{equation}\label{eq:Tjimmprime1plus2}
\widetilde{T}_{j|imm'}(x|y) = \mathbf{(1)}  + \mathbf{(2)},
\end{equation} 
where the first term corresponds to a jump at the end of the waiting-time on the node, whereas the second term is for the jump of a trapped walker. The computation of both terms requires first to determine the probability for an edge to be (un)available some time after having (not) jumped across it. 


\subsubsection{Corrections on  $p = \langle U \rangle/(\langle U \rangle + \langle D \rangle)$}
When the walker returns to a node after completion of a 2-cycle, the next destination node depends on the choice  previously made from the same location. First, the outgoing edge that was selected at the beginning of the cycle, say $i \rightarrow j$, has an increased probability (with respect to $ p = \langle U \rangle/(\langle U \rangle + \langle D \rangle)$) to still be available. The smaller the time $y = \tau-\nu$ to go through the cycle  and the subsequent walker's waiting-time, the more pronounced this effect.  Secondly, the converse is also true for any edge, say $i \rightarrow j'$, that wasn't selected. Not having been chosen in the past indicates a higher probability to have been and still be down some short time later.  In the main body, we present the derivation for the first effect, 
\begin{equation}\label{eq:pstar}
p_i^*(s,\nu) = \mathrm P\left\{ i \rightarrow j \mbox{ is up at  } s \, | \, \mbox{jumped across it at $\nu$}  \right\},
\end{equation}
whereas appendix~\ref{app:pdagger} contains the computations for the second effect quantified by
\begin{multline}\label{eq:pdagger}
p_i^\dagger(s,\nu) \\= \mathrm P\left\{ i \rightarrow j' \mbox{ is up at  } s \, | \, \mbox{jumped across $i \rightarrow j$ at $\nu$}  \right\}
\end{multline}
for some $s \geq \nu$ and $j'\neq j'$. Let us focus on the first effect, measured by the difference between $p^*(s,\nu)$ and $p$. Observe that this function only depends on the difference $s-\nu$.  Let us  define $\tilde{q_i}$, the probability that the jump $i \rightarrow j$ at time $\nu$ was done at the beginning of an up-time, that is to say, the walker was frustrated at the time of the jump. Observe that we do not know the effective waiting-time on the node before the jump (a longer waiting-time would have made a jump after a frustration period more plausible). Hence, assuming no memory beyond the last two jumps we have
\begin{equation}\label{eq:qi}
\tilde{q_i} = (1-p)^{|V_i|}. 
\end{equation}
Let us also define 
\begin{equation}\label{eq:U-tilde}
\widetilde{U}_i(x) = \tilde q_i U(x) + (1-\tilde q_i) \mathscr U(x), \quad x \geq 0,
\end{equation}
the density of the remaining  up-time of edge $i \rightarrow j$ after the jump at time $\nu$ was performed, where $\mathscr U$ is computed similarly to \eqref{eq:PDF-theta} : 
%\begin{equation}\label{eq:PDF-Umathscr}
$\mathscr U(x) = 1/\langle U \rangle \times \int_x^\infty U(x') \mathrm d x'.   $
%\end{equation}

\begin{rmk}
	The value of $\tilde{q}_i$ is irrelevant in \eqref{eq:U-tilde} if $U$ is an exponential density, because then $U = \mathscr U$. In that case, $\widetilde{U}_i$ does not depend on the strength of node $i$ in $\mathcal{ G}$ and we will drop the node-related index. 
\end{rmk}

As illustrated by  figure~\ref{fig:p-star}, we  can write
\begin{multline}\label{eq:pstarfollowingfigure}
p_i^*(s,\nu) = \int_{s-\nu}^\infty \widetilde{U}_i(r) \mathrm dr \\
+ \int_0^{s-\nu} \left(\widetilde{U}_i*D\right)(r) \int_{s-(\nu+r)}^\infty U(t) \, \mathrm d t \, \mathrm d r\\
+ \int_0^{s-\nu} \left(\widetilde{U}_i*D*U*D\right)(r) \int_{s-(\nu+r)}^\infty U(t) \, \mathrm d t  \, \mathrm d r+\ldots
\end{multline}
Introducing the notation for repeated convolutions
\begin{equation}\label{eq:notationforconvolution}
f^{*k_1} * g^{*k_2} = \underbrace{f*\dots *f}_{k_1 \mbox{ factors}} * \underbrace{g*\ldots*g}_{k_2 \mbox{ factors}}, \quad k_1,k_2 \in \mathbb{N},
\end{equation}
equation \eqref{eq:pstarfollowingfigure} has the  compact form
\begin{multline}\label{eq:pstar-final}
p_i^*(s,\nu) = \int_{s-\nu}^\infty \widetilde{U}_i(r) \mathrm dr \\
+ \sum_{k = 0}^{\infty} \int_0^{s-\nu} \left(\widetilde{U}_i*D^{*(k+1)}*U^{*k}\right)(r) \int_{s-(\nu+r)}^\infty U(t) \, \mathrm d t \, \mathrm d r. 
\end{multline}

% FIGURE : p^*
\begin{figure}[] 
	\centering
	\hspace*{-2em}
	\begin{center}
		\begin{tikzpicture}[scale = .8,transform shape]
		%\draw[help lines,line width=1,gray,thin] (-1, -1) grid (10,3);
		
		% AXIS FOR A_ij 
		
		\path[draw,fill = gray!2!white,draw = blue!15!white] (-1,-1.5) rectangle (10,2);
		\begin{scope}[yshift = -4cm]
		\path[draw,fill = gray!2!white,draw = blue!15!white] (-1,-1.5) rectangle (10,2);
		\end{scope}
		\begin{scope}[yshift = -8cm]
		\path[draw,fill = gray!2!white,draw = blue!15!white] (-1,-1.5) rectangle (10,2);
		\end{scope}
		
		\draw[ thick,->] (0,-.65)node[below]{$\nu$}--(0,0)node[anchor=south east]{down} 
		--(0,1)node[anchor=east]{up $\ $ }
		-- (0,1.5) node[anchor=south]{$i\rightarrow j$};
		\draw (-.15,0) -- (.15,0); % tick on vertical axis
		\draw(-.15,1)--(.15,1); % second tick on vertical axis
		\draw[thick,->](-.3,-0.5) -- (9.5,-.5)node[anchor = north](t){time\rule{0pt}{1.1em}};
		
		%\only<1>{%
			% EVOLUTION OF A_ij
			%\draw[fill] (0,0)circle(.1); 
			\draw[line width = 2.3pt](0,1)--(8.5,1)node[pos = .5,above]{$\widetilde{U}_i$}; %
			%(2,0)--(3,0)node[pos = .5,above]{$D$}  %
			%(3,1)--(4.5,1)node[pos = .5,above]{${U}$} % 
			%(4.5,0)--(6.7,0)node[pos = .5,above]{$D$} %
			%(6.7,1)--(8.5,1);
			
			% ADD GREEN COLOR
			%%\draw[line width = 2.3pt](0,1)--(2,1)node[pos = .5,above]{$\widetilde{U}$}; %
			%\draw[line width = 2.3pt,color = blue!55!green](2,0)--(3,0)node[pos = .5,above]{$D$}  %
			%(3,1)--(4.5,1)node[pos = .5,above]{${U}$} % 
			%(4.5,0)--(6.7,0)node[pos = .5,above]{$D$} ;%
			%%(6.7,1)--(8.5,1);

			% times
			\draw (6.7,-.35)--(6.7,-.65)node[below]{$\nu+r$}; 
			\draw[<->] (0,-1.2)--(6.7,-1.2)node[pos = .5,fill = white]{$r$};
			\draw (6.7,-.35)--(6.7,-.65)node[below]{$\nu+r$}; 
			\draw (8,-.35)--(8,-.65)node[below]{$s$};

		\begin{scope}[yshift = -4cm]
		
		% AXIS FOR A_ij 
		\draw[ thick,->] (0,-.65)node[below]{$\nu$}--(0,0)node[anchor=south east]{down} 
		--(0,1)node[anchor=east]{up $\ $ }
		-- (0,1.5) node[anchor=south]{$i\rightarrow j$};
		\draw (-.15,0) -- (.15,0); % tick on vertical axis
		\draw(-.15,1)--(.15,1); % second tick on vertical axis
		\draw[thick,->](-.3,-0.5) -- (9.5,-.5)node[anchor = north](t){time\rule{0pt}{1.1em}};
		
		%	\only<2>{%
		% EVOLUTION OF A_ij
		%\draw[fill] (0,0)circle(.1); 
		\draw[line width = 2.3pt](0,1)--(2,1)node[pos = .5,above]{$\widetilde{U}_i$}; %
		%(2,0)--(3,0)node[pos = .5,above]{$D$}  %
		%(3,1)--(4.5,1)node[pos = .5,above]{${U}$} % 
		%(4.5,0)--(6.7,0)node[pos = .5,above]{$D$} %
		%(6.7,1)--(8.5,1);
		
		% ADD GREEN COLOR
		%\draw[line width = 2.3pt](0,1)--(2,1)node[pos = .5,above]{$\widetilde{U}$}; %
		\draw[line width = 2.3pt,color = black](2,0)--(6.7,0) ;% continuous black
		\draw[line width = 2.3pt,color = blue!55!green,dashed](2,0)--(6.7,0)node[pos = .5,above]{$D$}  ;% green
		%(3,1)--(4.5,1)node[pos = .5,above]{${U}$} % 
		%(4.5,0)--(6.7,0)node[pos = .5,above]{$D$} ;%
		\draw[line width = 2.3pt](6.7,1)--(8.5,1);
		\node[style = {rectangle},rounded corners,draw,dotted,inner sep = 10pt,fill = white] (text) at (5.6,1.3) {\textcolor{black}{$k = 0$}};
		
		% times
		\draw (6.7,-.35)--(6.7,-.65)node[below]{$\nu+r$}; 
		\draw[<->] (0,-1.2)--(6.7,-1.2)node[pos = .5,fill = white]{$r$};
		\draw (6.7,-.35)--(6.7,-.65)node[below]{$\nu+r$}; 
		\draw (8,-.35)--(8,-.65)node[below]{$s$}; 
		\end{scope}
	
		\begin{scope}[yshift = -8cm]
		% AXIS FOR A_ij 
		\draw[ thick,->] (0,-.65)node[below]{$\nu$}--(0,0)node[anchor=south east]{down} 
		--(0,1)node[anchor=east]{up $\ $ }
		-- (0,1.5) node[anchor=south]{$i\rightarrow j$};
		\draw (-.15,0) -- (.15,0); % tick on vertical axis
		\draw(-.15,1)--(.15,1); % second tick on vertical axis
		\draw[thick,->](-.3,-0.5) -- (9.5,-.5)node[anchor = north](t){time\rule{0pt}{1.1em}};
		
		%\only<3>{%
		% EVOLUTION OF A_ij
		%\draw[fill] (0,0)circle(.1); 
		\draw[line width = 2.3pt](0,1)--(2,1)node[pos = .5,above]{$\widetilde{U}_i$} %
		(2,0)--(3,0)node[pos = .5,above]{$D$}  %
		(3,1)--(4.5,1)node[pos = .5,above]{${U}$} % 
		(4.5,0)--(6.7,0)node[pos = .5,above]{$D$} %
		(6.7,1)--(8.5,1);
		
		% ADD GREEN COLOR
		%\draw[line width = 2.3pt](0,1)--(2,1)node[pos = .5,above]{$\widetilde{U}$}; %
		\draw[line width = 2.3pt,color = blue!55!green,dashed](2,0)--(3,0)node[pos = .5,above]{$D$}  %
		(3,1)--(4.5,1)node[pos = .5,above]{${U}$} % 
		(4.5,0)--(6.7,0)node[pos = .5,above]{$D$} ;%
		%(6.7,1)--(8.5,1);
		\node[style = {rectangle},rounded corners,draw,dotted,inner sep = 10pt,fill = white] (text) at (5.6,1.3) {\textcolor{black}{$k = 1$}};
		%	}
		
		% times
		\draw (6.7,-.35)--(6.7,-.65)node[below]{$\nu+r$}; 
		\draw[<->] (0,-1.2)--(6.7,-1.2)node[pos = .5,fill = white]{$r$};
		\draw (6.7,-.35)--(6.7,-.65)node[below]{$\nu+r$}; 
		\draw (8,-.35)--(8,-.65)node[below]{$s$}; 
		\end{scope}
		\end{tikzpicture}
	\end{center}
	\caption{Parameters involved in the computation of ${p_i^*(s,\nu)}$. The schematics represent respectively the first, second and third terms  in equation \eqref{eq:pstar-final}. The three corresponding scenarios are the following. Either edge $i\rightarrow j$ remains up since time $\nu$ and up to time $s$, or it switches states twice before $s$, or it does so exactly four times on the interval $(\nu,s)$.  }
	\label{fig:p-star}
\end{figure}
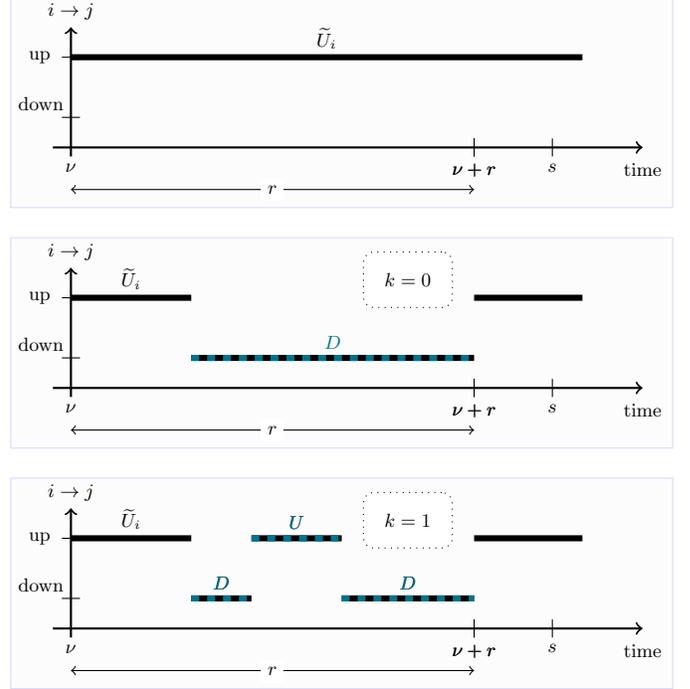
\begin{rmk}
In contrast with $p =  \frac{\langle U \rangle}{\langle U \rangle + \langle D \rangle}$, the expression for $p_i^*$ depends on the whole distribution of $U$, and not only on its mean. Also note that it only depends on the difference $s-\nu$, which is the time since the previous jump. See  figure~\ref{fig:pstar} for a numerical illustration in the all-exponential case. 
\end{rmk}

\subsubsection{The second case :  $ T_{j|iji}(t|\tau,\nu)$}
Having computed the necessary corrections on $p$, we are now in position to further develop equation~\eqref{eq:Tjimmprime1plus2}.  The first term - the walker is not trapped when he jumps -  reads
\begin{align}\label{eq:term1-corr}
\mathbf{(1)}_{(j|iji)} & = \psi_i(t-\tau) \sum_{k =1}^{|V_i|} \frac{1}{k} p_i^*(t,\nu) \binom{|V_i|-1}{k-1} \nonumber \\
                             & \rule{3em}{0pt}  \times (p_i^\dagger(t,\nu))^{k-1} (1-p_i^\dagger(t,\nu))^{|V_i|-k} \nonumber \\
				   & = \frac{p^*(t,\nu)}{p_i^\dagger(t,\nu)} \psi_i(t-\tau) \left[\frac{1-(1-p_i^\dagger(t,\nu))^{|V_i|}}{|V_i|}\right]. 
\end{align}
We notice that this expression is the same as for the acyclic graphs, up to a correction factor $p_i^*(t,\nu)/p_i^\dagger(t,\nu)$, and after  having replaced  $p$ by $p_i^\dagger(t,\nu)$.

\begin{figure}[] 
	\centering
	\hspace*{-.4em}
	\begin{tikzpicture}[scale = .77,transform shape]
	\path[draw,fill = gray!1!white,draw = blue!15!white] (-1.0,-1.5) rectangle (10,3.2);
	%\draw[help lines,line width=1,gray,thin] (-1, -1) grid (10,3);
	
	% AXIS FOR A_ij 
	\begin{scope}[yshift = 1cm]
	\draw[ thick,->] (0,-.65)node[below]{$\nu$}--(0,0)node[anchor=south east]{down} 
	--(0,1)node[anchor=east]{up}
	-- (0,1.5) node[anchor=south]{$i\rightarrow j$};
	\draw (-.15,0) -- (.15,0); % tick on vertical axis
	\draw(-.15,1)--(.15,1); % second tick on vertical axis
	\draw[thick,->](-.3,-0.5) -- (9.5,-.5)node[anchor = north](t){\rule{0pt}{1.1em}time};
	
	% EVOLUTION OF A_ij
	%\draw[fill] (0,0)circle(.1); 
	\draw[line width = 2.3pt](0,1)--(2,1)node[pos = .5,above]{$\widetilde{U}_i$} %
	(2,0)--(3,0)node[pos = .5,above]{$D$}  %
	(3,1)--(4.5,1)node[pos = .5,above]{${U}$} % 
	(4.5,0)--(7,0)node[pos = .5,above]{$D$} %
	(7,1)--(8.5,1);
	
	% times
	\draw (4.5,-.35)--(4.5,-.65)node[below]{$\nu+r$}; 
	\draw[<->] (0,-1.2)--(4.5,-1.2)node[pos = .5,fill = white]{$r$};
	\draw (6.1,-.35)--(6.1,-.65)node[below]{$s$}; 
	\draw (6.1,-1)node[below,text width = 1.3cm,fill = gray!10!white,rounded corners,draw,dotted]{\centering walker\\ready\\ to jump}; 
	\draw (7,-.35)--(7,-.65)node[below]{$t$}; 
	\end{scope}
	\end{tikzpicture}
	\caption{Parameters involved in the second term of ${T}_{jiji}(t|\tau,\nu)$ given by  equation \eqref{eq:term2-corr}. The figure corresponds to the term with $k  = 1$, that is to say the first up-time is followed by $k = 1$ down-up cycle. }
	\label{fig:term2-corr}
\end{figure}
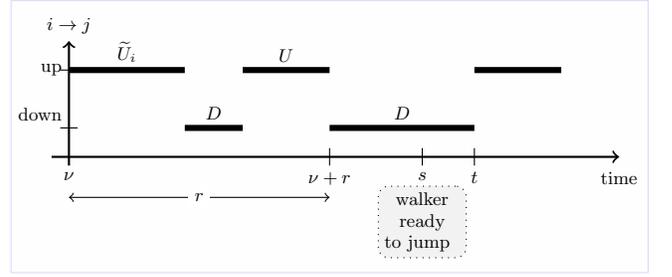

Using the same approach as for $p_i^*$, we obtain  the second term of  $ T_{j|iji}(t|\tau,\nu)$ corresponding to a trapped walker making the jump : 
\begin{multline}\label{eq:term2-corr}
\mathbf{(2)}_{(j|iji)}= \int_\tau^t \psi_i(s-\tau) \\
	\times \left[    \sum_{k = 0}^{\infty}    \int_0^{s-\nu} \left(      \widetilde U_i* D^{*k} * U^{*k}       \right) (r) \times D(t-\nu-r)\, \mathrm d r        \right] \\
	\times \left[    (1-p_i^\dagger(s,\nu)) \mathrm P \left\{     w>t-s         \right\}          \right]^{|V_i|-1} \, \mathrm d s. 
\end{multline}
The parameters are illustrated by figure~\ref{fig:term2-corr}. Relying on the previous computation of $p_i^*(s,\nu)$, expression~\eqref{eq:term2-corr} simplifies to the following one~: 
\begin{multline}\label{eq:term2-corrBIS}
\mathbf{(2)}_{(j|iji)}= \int_\tau^t \psi_i(s-\tau) 
(1-p_i^*(s,\nu)) \mathscr D (t-s)\\
\times \left[    (1-p_i^\dagger(s,\nu)) \mathrm P \left\{     w>t-s         \right\}          \right]^{|V_i|-1} \, \mathrm d s. 
\end{multline}
In this alternative form, $(1-p_i^*(s,\nu)) \mathscr D (t-s)$ refers to the probability that edge $i\rightarrow j$ is down at time $s$, and will remain so exactly until time $t$ when it becomes available to jumper again.

% FIGURE FOR SECOND CASE  : T_JIJI

\subsubsection{The third case :  $ T_{j|imi}(t|\tau,\nu)$ with $m\neq j$}
The first term of the transition density in the case of figure~\ref{fig:cycle_case3} is given by 
\begin{multline}\label{eq:term1-case3}
\mathbf{(1)}_{(j|imi)}=  \psi_i(t-\tau) \\
\times \Big [  p_i^*(t,\nu)   \times  \mathrm P \left\{  \mbox{choose } j \ | \ (i\rightarrow m) \mbox{ is up}    \right\} + \\
(1-p_i^*(t,\nu))  \times   \mathrm P \left\{  \mbox{choose } j \ | \ (i \rightarrow m) \mbox{ is down}    \right\}      \Big ]   ,
\end{multline}
where the two still undetermined probabilities are for events at time $t$. We can write
\begin{align}
 &\mathrm P \left\{  \mbox{choose } j \ | \ (i\rightarrow m) \mbox{ is up}    \right\} \nonumber \\ 
 &\quad  =  p_i^\dagger(t,\nu) \sum_{k = 0}^{|V_i|-2} \binom{|V_i|-2}{k}   \nonumber \\ 
 &\quad \qquad \qquad \qquad \times \frac{1}{k+2}     (p_i^\dagger(t,\nu))^k(1-p_i^\dagger(t,\nu))^{|V_i|-k-2}   \nonumber  \\
 &\quad  = \frac{|V_i|p_i^\dagger(t,\nu)+(1-p_i^\dagger(t,\nu))^{|V_i|-1}}{|V_i| (|V_i|-1)p_i^\dagger(t,\nu)}\label{eq:choosej_imup}
\end{align}
and
\begin{align}
&\mathrm P \left\{  \mbox{choose } j \ | \ (i \rightarrow m) \mbox{ is down}    \right\}   \nonumber \\
&\quad  = p_i^\dagger(t,\nu)\sum_{k = 0}^{|V_i|-2} \binom{|V_i|-2}{k}      \nonumber \\
&\quad \qquad \qquad \qquad \times \frac{1}{k+1}  (p_i^\dagger(t,\nu))^k(1-p_i^\dagger(t,\nu))^{|V_i|-k-2}  \nonumber \\
&\quad =  \frac{1-(1-p_i^\dagger(t,\nu))^{|V_i|-1}}{|V_i|-1} , \label{eq:choosej_imdown}
\end{align}
where  the final forms \eqref{eq:choosej_imup} and \eqref{eq:choosej_imdown} were obtained as in appendix~\ref{app:pdagger} using identity~\ref{eq:identitybinomials}. 

The second term can be shown to have the same expression as in \eqref{eq:term2-corrBIS}.

\subsection{The all-exponential case of table \ref{tab:limit-cases}}
We turn to the case where the three densities are exponential~:  $\psi$ has rate  $\mu$, $U$ has rate $\eta$ and $D$ has rate $\lambda$. Wherever possible, we drop the index of the node dependence, such that for instance $p_i^*$ becomes $p^*$. Let us recall that in this case, $\mathscr U  = U$ and  $\widetilde{U} = U$.  

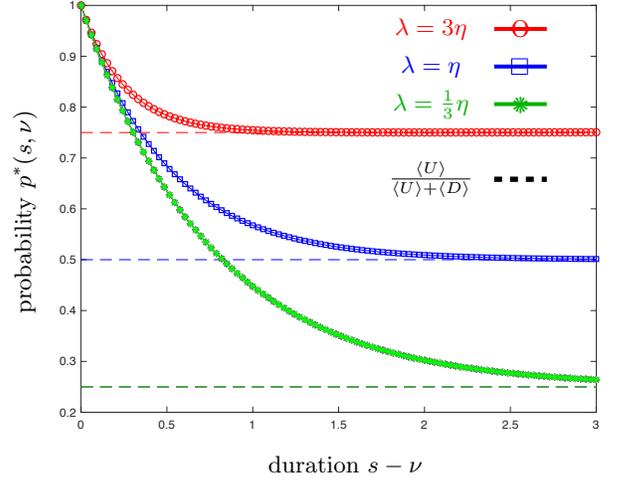
\begin{figure}
	\begin{tikzpicture}
	\node (fig) {\includegraphics[width = .85\columnwidth]{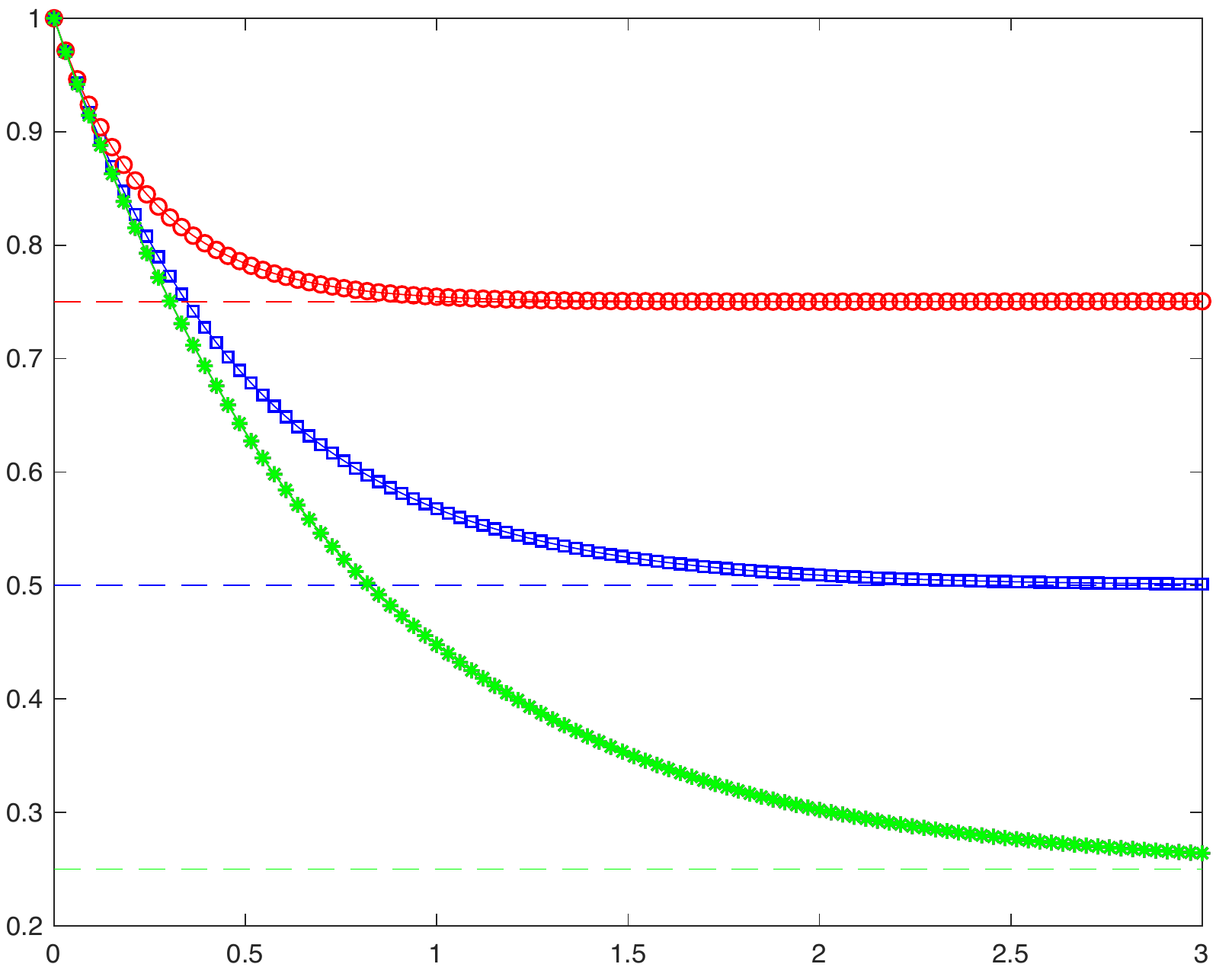}}; 
	\node[below = .1em of fig,xshift = .5em](xlablel){duration $s-\nu$}; 
	\node[above left = -1.5  and .1em of fig, rotate  = 90](ylablel){probability $p^*(s,\nu)$}; 
	%\node[above left = -2 and -14.3 em of fig](inset){\includegraphics[width= 0.35\columnwidth]{validationZ.pdf}};
	
	\begin{scope}[xshift = 1.3cm, yshift = 2.5cm,scale = .7]
	\node[color = red] at(0,0) (n2) {$\lambda = 3 \eta $}; 
	\draw[color = red, line width = 1.5pt] (1.2,0)--(2.2,0) node[pos = .5]{\Large o}; 
	\end{scope}
	
	\begin{scope}[xshift = 1.3cm,yshift = 2cm,scale = .7]
	\node[color = blue] at(0,0) (n1) {$\lambda = \eta $}; 
	\draw[color = blue, line width = 1.5pt] (1.2,0)--(2.2,0) node[pos = .25]{\qed}; 
	\end{scope}

	\begin{scope}[xshift = 1.3cm,yshift = 1.5cm,scale = .7]
	\node[color = green!70!black] at(0,0) (n3) {$\lambda = \frac{1}{3} \eta $}; 
	\draw[color = green!70!black, line width = 1.5pt] (1.2,0)--(2.2,0) node[pos = .5,yshift = -.05em]{\Large $*$}node[pos = .5,yshift = -.05em,rotate = 30]{\Large $*$};  
	\end{scope}
	
	\begin{scope}[xshift = 1.3cm,yshift = 0.5cm,scale = .7]
	\node[color = black] at(0,0) (n3) {$\frac{\langle U \rangle}{\langle U \rangle + \langle D \rangle}$}; 
	\draw[color = black, line width = 2pt, dashed] (1.2,0)--(2.2,0); 
	\end{scope}
	
	%\draw[xshift = 1cm] (-.6,1.25) rectangle (2.3,2.7); 
	\end{tikzpicture}
	
	\caption{Evolution of $p^*(s,\nu)$, i.e. the probability for an edge to be in the up-state at time $s$ knowing it was available at time $\nu$, in the all-exponential case for various ratios of $\eta/\lambda$. The red series with circle markers and the green one with star markers come from equation~\eqref{eq:pstar-final-allexpo}, whereas the blue series in the middle with square markers corresponds to~\eqref{eq:pstar-both-expo}.  In all three cases, the dotted lines are the corresponding values of $p =  \langle U \rangle/(\langle U \rangle + \langle D \rangle)$ that assume no prior information. }
	\label{fig:pstar}
\end{figure}

The expression of $p^*$ given in  \eqref{eq:pstar-final} and the second term of the transition density given in \eqref{eq:term2-corr} both require to compute the density $U^{*k} * D^{*k}$, which corresponds to the sum of the random variables $X_U^{(k)}  + X_D^{(k)}$ where $X_U^{(k)}$ (resp. $X_D^{(k)}$) is the sum of $k$ exponential random variables with parameter $\eta$ (resp. $\lambda$). It is well known that $X_U^{(k)} \sim \mathrm{Erlang}(k,\eta)$, and $X_D^{(k)} \sim \mathrm{Erlang}(k,\lambda)$. Using \cite{jasiulewicz2003convolutions} for the convolution of Erlang densities, we find that the density of $X_U^{(k)} + X_D^{(k)}$ is given by 
\begin{multline}\label{eq:density-sum-erlang}
	f_{X_U^{(k)} + X_D^{(k)}}(t) = \frac{(\eta \lambda)^k}{(\lambda-\eta)^{2k}}    \\
		\times  \sum_{j=1}^{k} \left[
		\frac{(-1)^{k-j}}{(j-1)!}     \binom{2k-j-1}{k-j} (\lambda - \eta )^j 
		\left\{     e^{-\eta t}  + (-1)^{j}  e^{-\lambda t}      \right\}
		\right]\\
		\times t^{j-1} \mathbbm{1}_{\mathbb{R}^+}(t). \\
	\end{multline}
It follows that \eqref{eq:pstar-final} becomes
\begin{multline}\label{eq:pstar-final-allexpo}
	p^*(s,\nu) = \int_{s-\nu}^\infty {U}(r) \mathrm dr \\
	+ \sum_{k = 0}^{\infty} \int_0^{s-\nu} f_{X_U^{(k+1)} + X_D^{(k+1)}}(r) \int_{s-(\nu+r)}^\infty U(t) \, \mathrm d t \, \mathrm d r. 
\end{multline}
Note again that  index $i$ is now needless. The above series can be truncated to allow for a practical computation. In the case that $U$ and $D$  share the same rate parameter $\lambda$, this expression further simplifies. A direct computation yields 
\begin{equation}\label{eq:pstar-both-expo}
p^*(s,\nu) = e^{-\lambda (s-\nu)} \cosh(\lambda(s-\nu)) = \frac{1}{2} \left( 1+ e^{-2\lambda(s-\nu)} \right). 
\end{equation}
The second term being positive is the increase with respect to $p = \frac{1}{2}$, and it is smaller for a higher rate $\lambda$ and for larger $s-\nu$. This is because more up/down cycles will decrease the memory effect on the state of the edge. A numerical illustration of \eqref{eq:pstar-final-allexpo} and \eqref{eq:pstar-both-expo} is offered by figure~\ref{fig:pstar}. 

\begin{figure}
	\centering
		\begin{tikzpicture}
	\node (fig) {\includegraphics[width = .85\columnwidth]{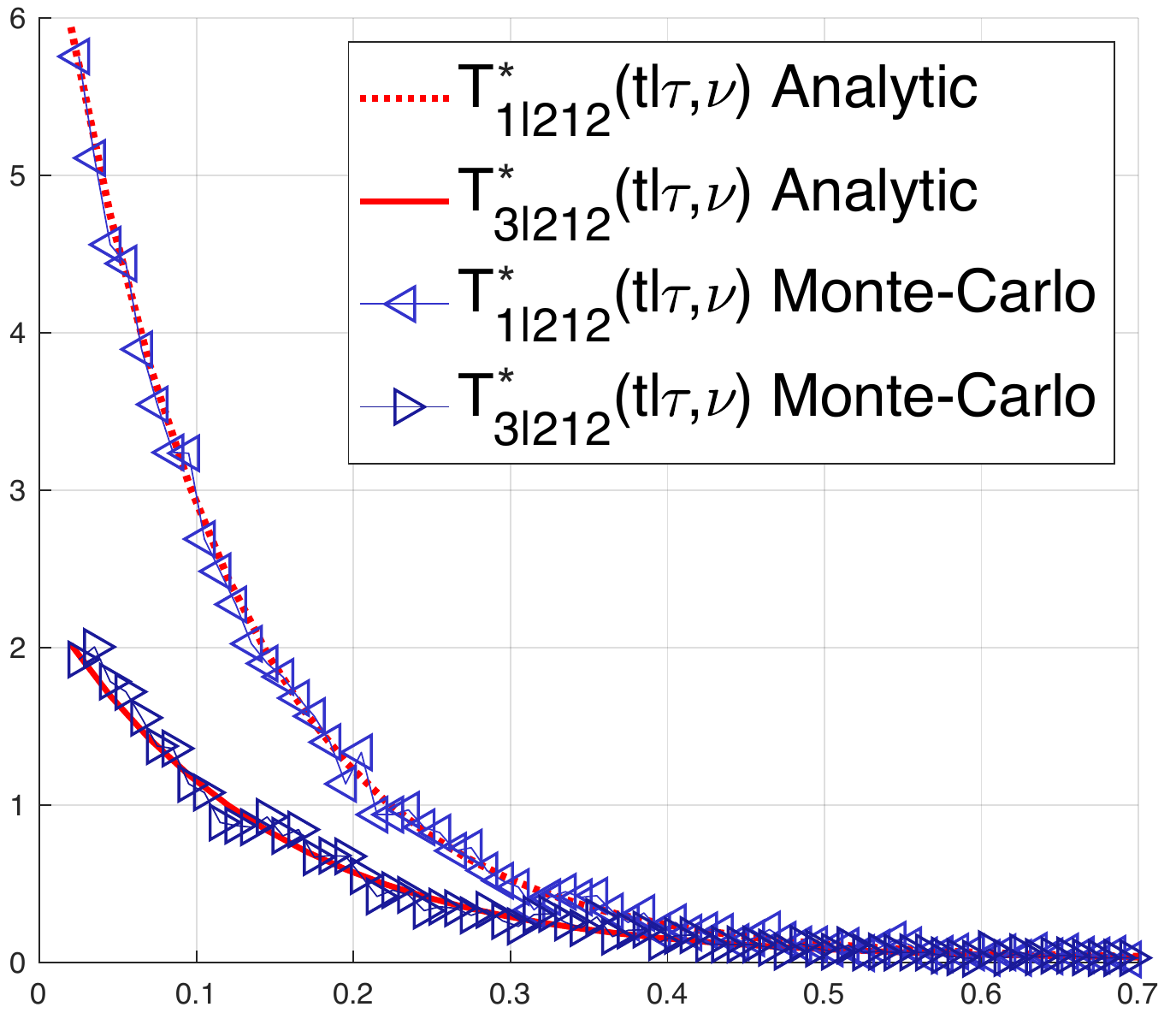}}; 
	\node[below = .1em of fig,xshift = .5em](xlablel){time $t$}; 
	\node[above left = -1.8  and .1em of fig, rotate  = 90](ylablel){transition density}; 
	\node at (-.5,2.75) {$*$};
	\node at (-.5,2.1) {$*$};
	\node at (-.5,1.46) {$*$};
	\node at (-.5,.835) {$*$};

	\begin{scope}[scale  = .8, transform shape, xshift = 2.5em, yshift = 4.4cm]
\node[style = circle, draw] (n1) at (0,0) {1};
\node[style = circle, draw] (n2) at (1.5,0) {2};
\node[style = circle, draw] (n3) at (3,0) {3};
\draw[very thick,->] (n1) to[bend left]  (n2); 
\draw[very thick,->] (n2) to[bend left]  (n1); 
\draw[very thick,->] (n2) to  (n3);
%\draw[very thick,->] (n3) to[bend right]  (n1);
\end{scope}

	\end{tikzpicture}
	\caption{Validation of the analytical formula for the memory effect related to $p^*(\cdot,\cdot)$  in the conditional transition density. The simultaneous effect of $p_i^\dagger(\cdot,\cdot)$ was annihilated by replacing it by $p$ in the formulas of the density, which is therefore written with the superscript $*$ in the legend. The Monte-Carlo simulation was designed so as to allow a memory effect solely on edge $2 \rightarrow 1$ of the graph appearing as an inset, thereby neglecting the effect corresponding to $p_i^\dagger$. The rate of the walker is  $\mu = 8$, the edges are characterized by the rates  $\lambda = 1 = \eta $, and $\tau = 0.02 = 2 \nu$. }
	\label{fig:TransitionVSmontecarlo}
\end{figure}
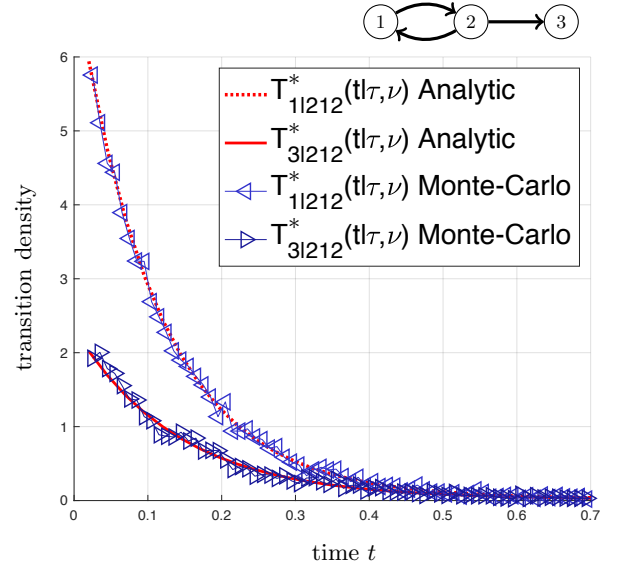

On figure~\ref{fig:TransitionVSmontecarlo} the correctness of the first correction by $p^*$ on $p$ is assessed through comparison with a Monte-Carlo simulation. In order to evaluate it independently from the concurrent correction due to $p_i^\dagger$, we have set $p_i^\dagger(\cdot,\cdot) = p$ in the formulas of the conditional transition density, which is then written as $T^*_{j|imm'}(t|\tau,\nu)$ to highlight the change. 

Let us consider  the second correction on $p = \langle U \rangle/(\langle U \rangle + \langle D \rangle)$, which is quantified by $p_i^\dagger$. Assuming again the same rate for $U$ and $D$, it follows directly from equations~\eqref{eq:pitildefinal} and \eqref{eq:pdagger_final} that 
\begin{equation}\label{eq:pdagger-both-exponential}
p_i^\dagger(s,\nu) = \frac{1}{2} - \frac{1}{4}e^{-2\lambda(s-\nu)}
\end{equation}
when we set $|V_i| = 2$, a choice that maximizes the importance of this effect. The  second term represents the difference with respect to $p = \frac{1}{2} = 2 \tilde{q_i}$, and is such that $p_i^\dagger(s,\nu) \rightarrow \tilde{q_i}  $ if $s-\nu \rightarrow 0^+$ and $p_i^\dagger(s,\nu) \rightarrow p $ if $s-\nu \rightarrow + \infty$.

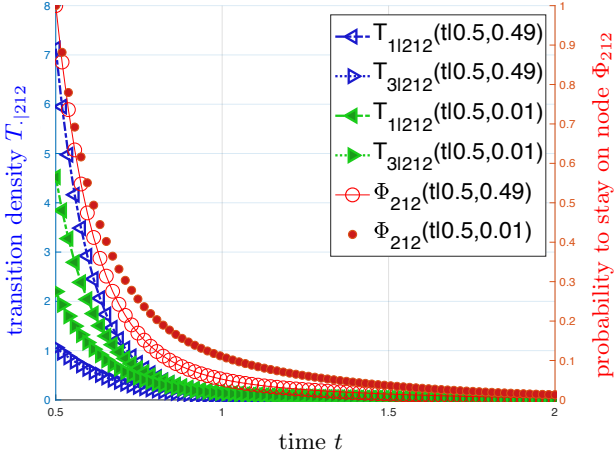
\begin{figure}
	\centering
	\hspace*{-.5em}
	\begin{tikzpicture}
	\node (f1)  {\includegraphics[width=.83\columnwidth]{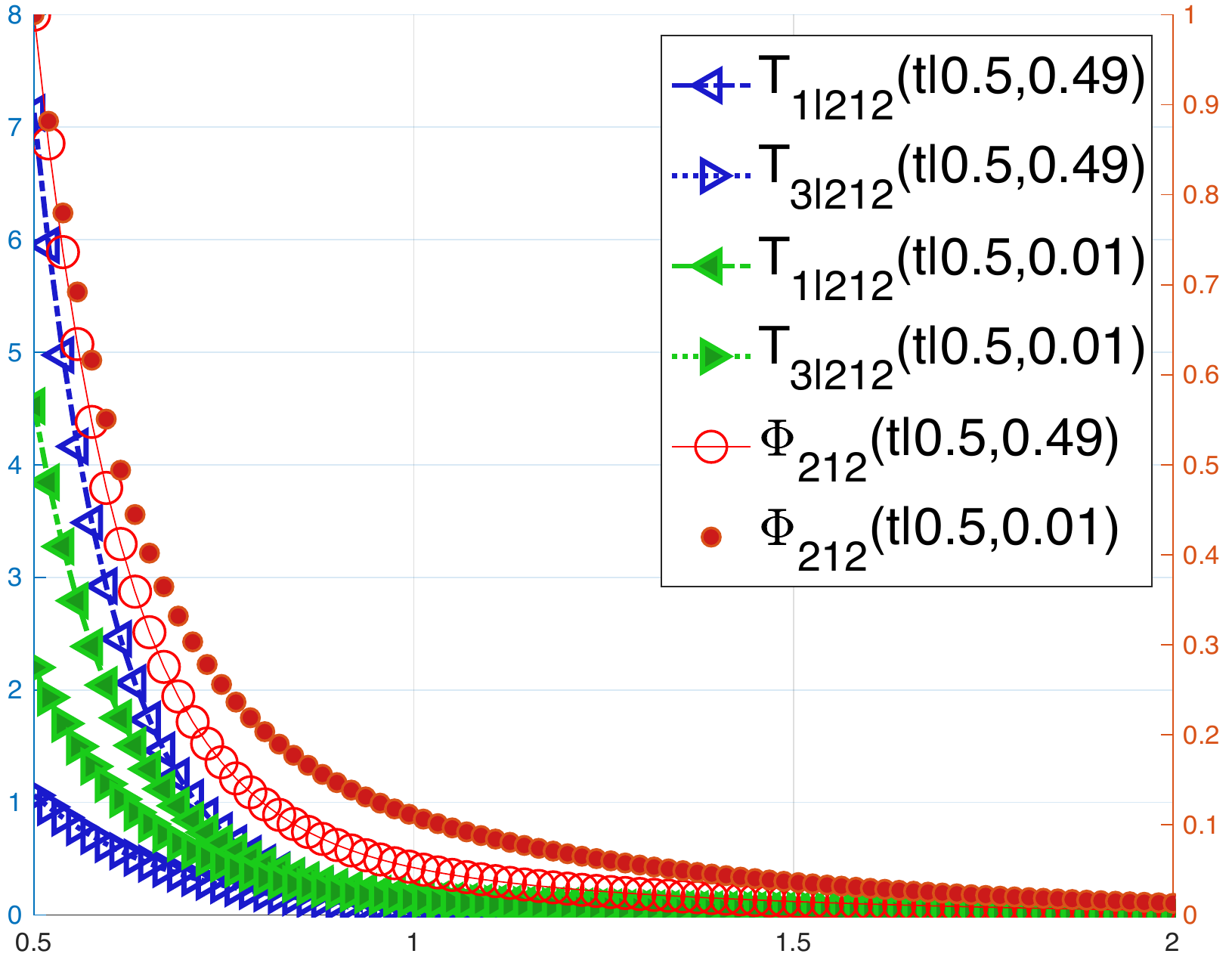}};
	\node[below = -.2em of f1] (time)   {time $t$};
	\node[rotate = 90, yshift = 3.9cm] (blue) {\textcolor{blue}{transition density $T_{\cdot | 212}$}};
	\node[xshift = .97\columnwidth,yshift = -1em,rotate = 90] (red) at (-4.5,.4) {\textcolor{red}{probability to stay on node $\Phi_{212}$}};
	\end{tikzpicture}
	
	\caption{Stronger (blue triangle markers) vs weaker (filled green markers) memory effect depending on the time to go through a cycle. On the left vertical axis, one sees that the differentiation between the jump densities towards nodes~1 and~3 respectively, is more pronounced when the duration $\tau-\nu$ is smaller, and decreases with $t$. The resulting probabilities to stay put on node~2 are plotted in red on the right vertical axis. The empty circle markers correspond to a strong memory effect, and indicate a lower probability to remain for a long time on the node before a jump, when compared to the series with filled red circle markers (weaker memory). The graph is the one of figure~\ref{fig:TransitionVSmontecarlo}. The rates are $\mu = 8$, $\eta = 1 = \lambda$, and $\tau = 0.5, \, \nu = 0.49$ for the strong effect, whereas $\tau = 0.5, \, \nu = 0.01$ in the other case. }
	\label{fig:increasedMemory}
\end{figure}

Combining the effects of $p^*$ and $p_i^\dagger$ results in figure~\ref{fig:increasedMemory} where it appears clearly that a shorter time to go around the cycle $2\rightarrow 1 \rightarrow 2$ induces a stronger bias in favor of another jump along $2 \rightarrow 1$ instead of $2 \rightarrow 3$. 

\begin{figure}
\hspace*{-1em}
\begin{tikzpicture}
\node (f1) at (0,0) {\includegraphics[width=.99\columnwidth]{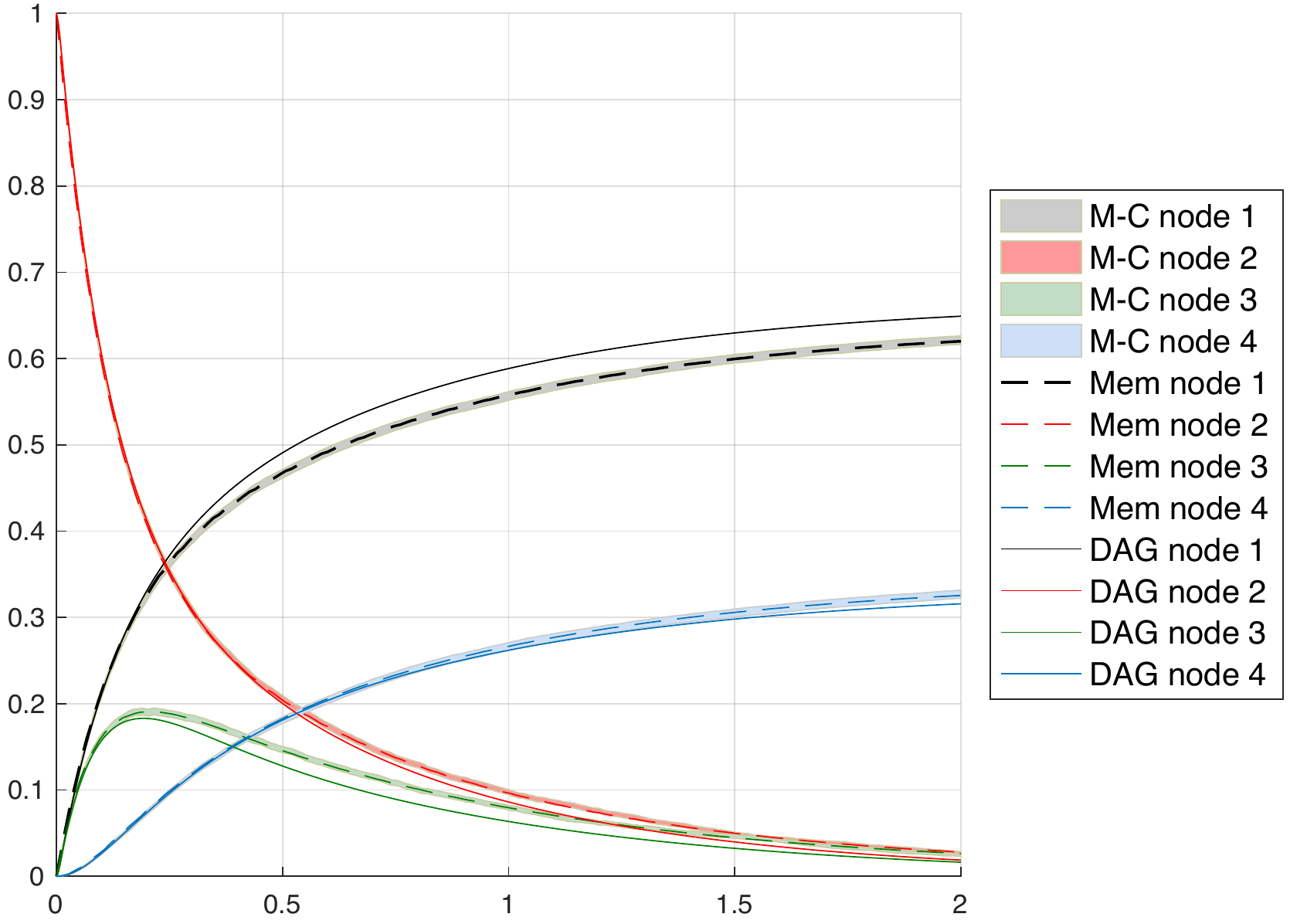}};
\node (time) at (-.8,-3.2) {time $t$};
\node[rotate = 90] (ni) at (-4.5,.4) {probability $n_i(t)$};

\begin{scope}[yshift = 1.1cm]
\path[draw,fill = white] 
(1.44,3.5)  node[style = circle,draw,text width = .4em,align = center](c1){$1$}
++(.85,0)  node[style =circle,draw,text width = .4em,align = center](c2){$2$}
++(.85,0)  node[style = circle,draw,text width = .4em,align = center](c3){$3$}
++(.85,0) node[style = circle,draw,text width = .4em,align = center](c4){$4$};
\end{scope}

\draw[->,very thick] (c2) -- (c1); 
\draw[->,very thick] (c2) to[bend left] (c3); 
\draw[->,very thick] (c3) to[bend left] (c2); 
\draw[->,very thick] (c3) -- (c4);

%\node[draw,fill = white] (n1) at (-.9,2) {\includegraphics[height =1.6cm]{shadedn1}};
\node[draw,fill = white,anchor = west] (n1) at (-4.55,4.6) {\includegraphics[height =1.6cm]{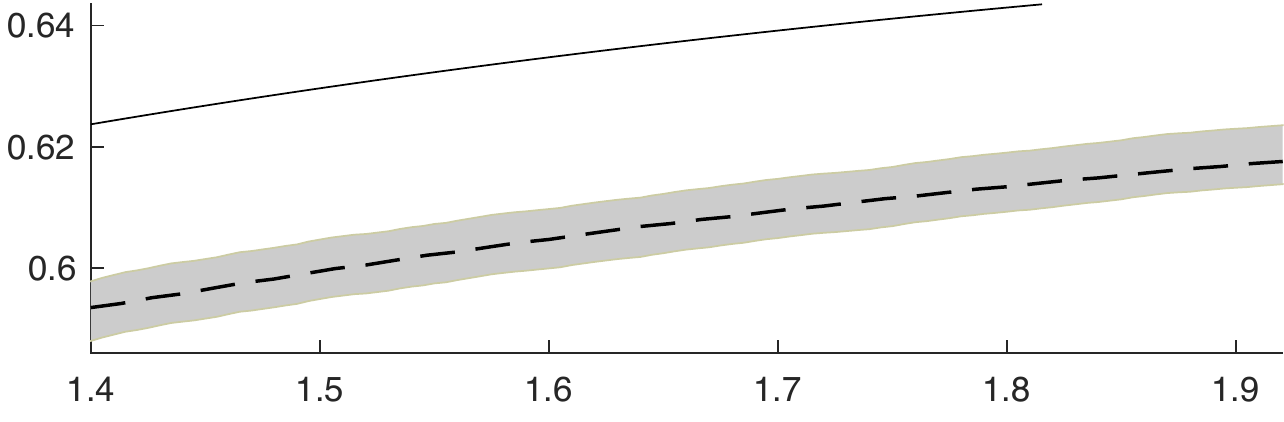}};
\path (-4.55,-5) node[anchor = west,draw] (n4) {\includegraphics[height=1.8cm]{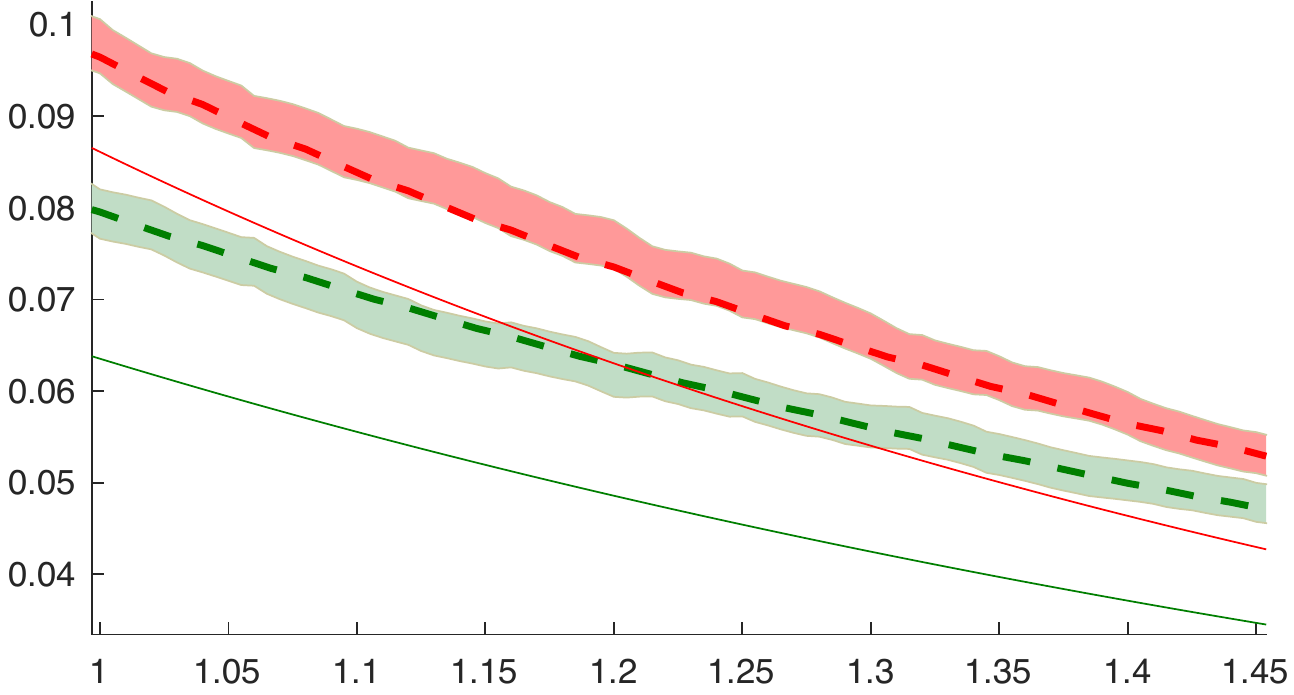}} ++(3.85,0)  node[draw,anchor = west] (n23) {\includegraphics[height=1.8cm]{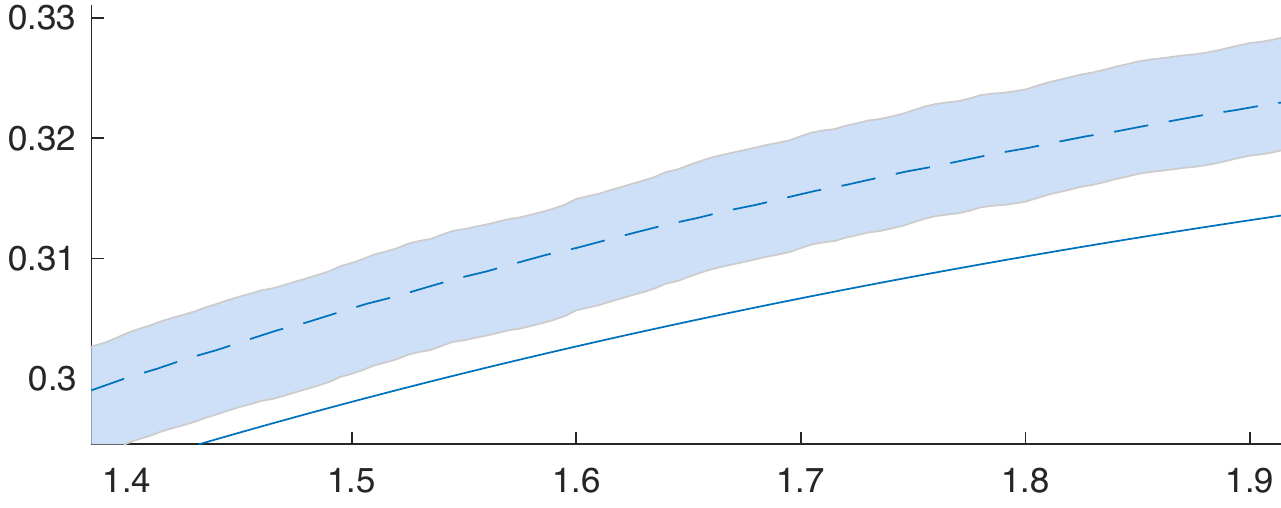}}; 

\node (a) at (0,5.2){(a)};
\node (b) at (-1.5,-4.6){(b)};		
\node (c) at (.4,-4.6){(c)};		

\node (aa) at (1.7,0.56){{\scriptsize (a)}};
\node (cc) at (1.7,-1.17){{\scriptsize (c)}};		
\node (bb) at (0.28,-2.25){{\scriptsize (b)}};

% draw rectangles
\draw[densely dotted, thick] (0.45,0.4) rectangle(2,1);
\begin{scope}[yshift = -1.71cm]
\draw[densely dotted,thick] (0.45,0.4) rectangle(2,1);
\end{scope}

\draw[densely dotted,thick] (-0.8,-2.6) rectangle(0.55,-2.1);

\draw[->,gray!60!white] (0.45,1) to (-1,3.6); 
\draw[->,gray!60!white] (2,-1.31) to[bend left=5] (2.5,-3.85); 
\draw[->,gray!60!white] (-0.8,-2.6) to[bend right=20] (-2.5,-3.85); 

%\draw[step=.5cm,red!14!white] (-4.4,-4.4) grid (4.4,4.4);
%\foreach \x in {-4,-3.5,...,4}
%\draw (\x cm,1pt) -- (\x cm,-1pt) node[anchor=north] {{\tiny $\x$}};
%\foreach \y in {-4,-3.5,...,4}
%\draw (1pt,\y cm) -- (-1pt,\y cm) node[anchor=east] {{\tiny $\y$}};
\end{tikzpicture}
\caption{Numerical validation of the analytical framework (dashed lines) accounting for the last-2-cycle memory effect. The Monte-Carlo simulation (shading) results from the average of $4 \cdot 10^4$ independent trajectories of a single walker. The shaded areas  determine an interval  centered around the mean, of width equal to twice the standard deviation. In this simulation, the walker always starts in node~2. Due to the cycles effect, the increase of $n_i(t)$ for node~1 (inset $(a)$) is much slower when compared with the curve resulting from the transition densities valid for acyclic graphs (solid lines). Indeed, the  memory effect comes into play only after (and if) the walker has completed the sequence $2 \rightarrow 3 \rightarrow 2$. This effect then acts in favor of node~4, for which the difference between the actual probability and the DAG approximation is less dramatic (inset $(c)$). Also observe that the memory effect tends to bring the curves corresponding to the two nodes belonging to the cycle closer closer together (nodes~2 and~3, inset $(b)$). By the same mechanism, the convergence of $n_2(t)$ and $n_3(t)$ to $0$ is notably slower. 	The dashed series resulting from the analytical modeling with corrections are virtually indistinguishable from the Monte-Carlo ones, which shows the effectiveness of the developed framework. The rates are $\mu = 8$, $\eta = 1 = \lambda$. }
\label{fig:validationCycles}
\end{figure}
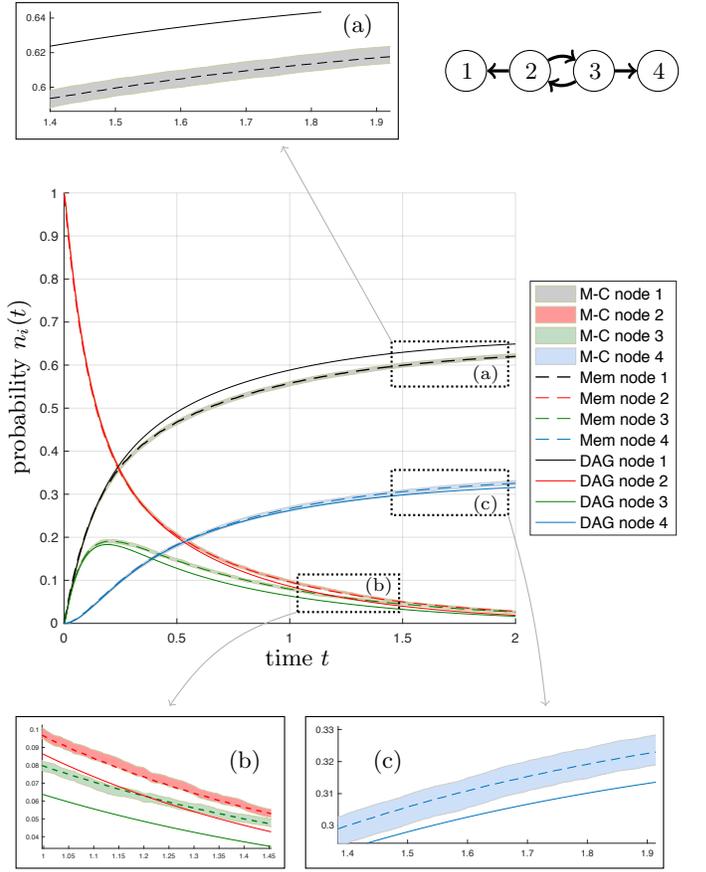

A validation of the comprehensive analytical framework through a simple numerical example is the purpose of figure~\ref{fig:validationCycles}. 

\section{Numerical methods}\label{sec:simulation}

We solved the Volterra vector integral equations \eqref{eq:volterraII} and \eqref{eq:q_tuple3} by applying a trapezoidal scheme for discretization of the integrals, by a method described in  \cite{delves1988computational}. The initial condition $\mathbf q^{(0)}(t) =  \mathbf{n}(0)\delta (t)$ arising in these equations was approximated using a half-gaussian-like positive function $\delta_\epsilon (t)$ parametrized by  a small parameter $\epsilon$, such that 
\begin{equation}\label{eq:smoothingDirac}
\mathbf q^{(0)} (t) \approx \mathbf n (0) \delta_\epsilon(t), \quad \int_0^\infty \delta_\epsilon (t) dt = 1. 
\end{equation}
The numerical method uses Monte-Carlo simulation to determine the  probabilities $\mathbf{n}(t)$ by averaging over a large set of realizations. Each trajectory of the walker corresponds to a new realization of the walker waiting-times and of the up- and down-time of the edges.  The time interval $[0,T)$ of the simulation is discretized according to some partition $0  = t_0 <t_1 < \dots  < t_m= T$. The probability for the walker to be in some node over some time window $[t_k,t_{k+1}]$ is approximated by the mean over all simulations, of the fraction of time spent by the walker on that  particular node. This is the same method as   in~\cite{hoffmann2012generalized}. 

\section{Conclusion}\label{sec:applications}
A very common assumption in the study of dynamical processes on networks is to take only the direction of the edges and  their weights into account. Accordingly, one often assumes that temporal events on the edges occur as a Poisson process. An important contribution of the field of temporal networks  is to question this assumption and to propose more complex temporal models, including renewal processes with arbitrary event-time distributions. Yet, in a majority of works, one considers, implicitly or explicitly, instantaneous interactions. The main purpose of this work was to incorporate edge duration in stochastic model of temporal networks, and to estimate its impact on random walk processes. We have derived analytical expressions for various properties of the process. As we have shown, those are exact on DAGs, and we have presented corrections due to the presence of cycles on the underlying network. 

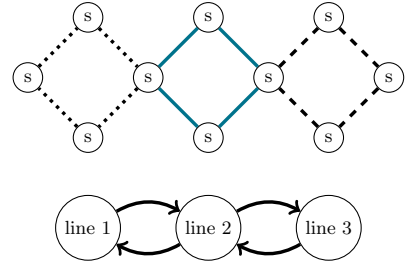
\begin{figure}
	\begin{center}
		\begin{tikzpicture}[scale = .8, transform shape]
		%\draw[help lines](0,0) grid (6,2); 
		\node[style = circle,draw] at(0,1)(s1){s}; 
		\node[style = circle,draw] at(1,2)(s2){s};
		\node[style = circle,draw] at(2,1)(s3){s};
		\node[style = circle,draw] at(1,0)(s4){s};
		\node[style = circle,draw] at(3,2)(s5){s};
		\node[style = circle,draw] at(4,1)(s6){s};
		\node[style = circle,draw] at(3,0)(s7){s};
		\node[style = circle,draw] at(5,2)(s8){s};
		\node[style = circle,draw] at(6,1)(s9){s};
		\node[style = circle,draw] at(5,0)(s10){s};
		
		\draw[dotted,line width = 1.3pt] (s1)--(s2)--(s3)--(s4)--(s1);
		\draw[blue!55!green,line width = 1.3pt] (s3)--(s5)--(s6)--(s7)--(s3);
		\draw[dashed,line width = 1.3pt] (s6)--(s8)--(s9)--(s10)--(s6);
		
		\begin{scope}[yshift = -1.5cm]
		\node[style = circle,draw] at (1,0) (l1) {line 1};
		\node[style = circle,draw] at (3,0) (l2) {line 2};
		\node[style = circle,draw] at (5,0) (l3) {line 3};
		\draw[line width = 1.3pt,->] (l1) to[bend left] (l2); 
		\draw[line width = 1.3pt,->] (l2) to[bend left] (l3); 
		\draw[line width = 1.3pt,->] (l2) to[bend left] (l1); 
		\draw[line width = 1.3pt,->] (l3) to[bend left] (l2); 
		\end{scope}
		\end{tikzpicture}
	\end{center}
\caption{A disruption-tolerant network based on mobile wireless sensors. The top schematic represents three independent bus lines, where two busstops are shared by two different lines. The underlying network of allowed connections is given in the bottom graph. }
\label{fig:busline}
\end{figure}

This work is mostly theoretical but it has plenty of potential applications in real-life systems. Take contact networks and their impact on epidemic or information spreading as a canonical example. In engineering, practical applications include peer-to-peer  and proximity networks of mobile sensors with wireless connections (cast under the framework of DTN : disruption / tolerant networks). A good example would be the diffusion of buses in a city that can communicate only when they halt at the same bus stop  \cite{figueiredo2012characterizing} (see   figure~\ref{fig:busline}).  Given the central role of random walks in the design of algorithms on networks, our results also open the way to generalise standard tools such as Pagerank for centrality measures and Markov stability for community detection \cite{Delvenne2010PNAS}.
Yet, in our view, the key message of this paper is its emphasis on the 
importance of three timescales to characterise diffusion on temporal networks, one for diffusion and two for the edge dynamics. Future research directions include a more thorough investigation on when certain timescales can be neglected over other ones, hence leading to simplified mathematical models, and models including a fourth timescale, associated to the possible non-stationarity of the network evolution, for instance due to circadian rhythms.

\appendix
\section{Transition density for DAGs  in case 2 of table~\ref{tab:limit-cases}}
\label{app:case2}
When link activation is instantaneous, $\langle U \rangle = 0$, $p=0$ and the  first term of $T_{ji}(t,\tau)$ vanishes. The second term yields 
\begin{equation}
T_{ji}(t,\tau) = \mu \lambda e^{\mu \tau - \lambda |V_i| t} \int_\tau^t e^{(-\mu + \lambda |V_i|)x} \mathrm d x
\end{equation}
If $\mu = \lambda |V_i|$, the integral equals  $t-\tau$ and $T_{ji}(t,\tau)  =  \lambda \mu (t-\tau) e^{-\mu (t-\tau)}$. Otherwise, a direct calculation yields
\begin{equation}\label{eq:Tjicase2final}
T_{ji}(t,\tau) = \frac{\lambda \mu}{\lambda |V_i|-\mu} \left(          e^{-\mu (t-\tau)} - e^{-\lambda |V_i| (t-\tau)}        \right). 
\end{equation}
Observe that taking the limit $\mu \rightarrow \infty $ in the above expression yields 
\begin{equation}
T_{ji}(t,\tau) = \frac{1}{|V_i|} \cdot \lambda |V_i| e^{-\lambda |V_i| (t-\tau)},
\end{equation}
where the second factor is the density of the minimum of $|V_i|$ independent exponential densities  with rate $\lambda$. We have recovered case~1. Starting from~\eqref{eq:Tjicase2final} we have 
\begin{equation}\label{eq:Phicase2final}
\Phi_i(t,\tau) = \frac{1}{\lambda |V_i| - \mu} \left(         \lambda |V_i| e^{-\mu (t-\tau)} - \mu e^{- \lambda |V_i| (t-\tau)}        \right). 
\end{equation}
The case that $\mu = \lambda |V_i|$ is straightforward. 

\section{Transition density for DAGs in case~3 of table~\ref{tab:limit-cases}}
\label{app:case3}
When  the link activation follows an exponential density $\mathcal E (\eta)$, we have $p = \frac{\lambda}{\lambda + \eta}$  and the first term of $T_{ji}(t,\tau) = \mathbf{(1)} + \mathbf{(2)} $ reads
\begin{equation}
\mathbf{(1)} = \mu e^{-\mu (t-\tau)} \frac{1}{|V_i|} \left(        1-(1-p)^{|V_i|}     \right) , 
\end{equation}
whereas the second term $\mathbf{(2)}$  in the more general case that $ \mu \neq \lambda |V_i|$  is given by~\eqref{eq:Tjicase2final} multiplied by $(1-p)^{|V_i|}$. Following a direct calculation, the probability to stay on node $i$ for a time of at least $t-\tau$ now reads
\begin{multline}
\Phi_i(t,\tau) = 1- (1-p)^{|V_i|} \\
\times  \left(     1 -        \frac{1}{\lambda |V_i| - \mu} \left(         \lambda |V_i| e^{-\mu (t-\tau)} - \mu e^{- \lambda |V_i| (t-\tau)}        \right)         \right )  \\
- 
 \left(        1-(1-p)^{|V_i|}     \right) \left(        1-e^{-\mu (t-\tau)}         \right).
\end{multline}

\section{Computation of $p_i^\dagger(s,\nu)$}
\label{app:pdagger}
We consider a two cycle $i \rightarrow j \rightarrow i$  of the underlying graph $\mathcal G $ where node $i$ has at least one neighbor $j'$ other than $j$. For the sake of compactness, we compute $p^\dagger(s,\nu)$ - the probability that edge $i \rightarrow j'$ is down at time $s$ knowing it wasn't selected by the walker at time $\nu$ in the past - under the assumption that the durations $U(t)$ and $D(t)$ follow the same distribution. The reasoning readily applies without this assumption. 

Let $E_s$ and $E_\nu$ denote respectively the events that $i\rightarrow j$ is up at time $s$ and at time $\nu$. Let $E'_s$ and $E'_\nu$ be the corresponding events for edge $i \rightarrow j'$ and let also $F_\nu$ be the event that the walker jumped through $i\rightarrow j$ at time $\nu$. We write $\bar{A}$ the complement of event $A$, such that $\mathrm{P} \left\{  A \cup \bar{A}   \right\} = 1$ and $\mathrm{P} \left\{  A \cap \bar{A}   \right\} = 0$. Using the law of total probabilities for conditional probabilities we have
\begin{align}\label{eq:pdag_step1}
p_i^\dagger (s,\nu) &= \mathrm{P} \left\{   E'_s \vert F_\nu    \right\}   \nonumber  \\
                               &= \mathrm{P} \left\{  E'_s \cap E'_\nu \vert F_\nu\right\} +   \mathrm{P} \left\{  E'_s \cap \overline{E'_\nu} \vert F_\nu\right\}   \nonumber  \\
                               &= \mathrm{P} \left\{  E'_s   \vert E'_\nu \cap F_\nu\right\}  \mathrm{P} \left\{   E'_\nu \vert F_\nu \right\}  \nonumber  \\   
                               & \qquad \qquad + \mathrm{P} \left\{  E'_s   \vert \overline{E'_\nu} \cap F_\nu\right\}  \mathrm{P} \left\{   \overline{E'_\nu} \vert F_\nu \right\}. 
\end{align}
Now, using the assumption that the up- and down-times follow the same distribution, $\mathrm{P} \left\{  E'_s   \vert E'_\nu \cap F_\nu\right\} = p_i^*(s,\nu)$ and $\mathrm{P} \left\{  E'_s   \vert \overline{E'_\nu} \cap F_\nu\right\}  = 1- p_i^*(s,\nu)$. Also observe that  $ \mathrm{P} \left\{   \overline{E'_\nu} \vert F_\nu \right\}  = 1-  \mathrm{P} \left\{   E'_\nu \vert F_\nu \right\} $. So it only remains to compute 
\begin{equation}\label{eq:pitilde}
\tilde p_i :=  \mathrm{P} \left\{   E'_\nu \vert F_\nu \right\}, 
\end{equation}
the probability for an edge to be available at some time, knowing a jump was performed through a competing edge at that time. This would yield the final expression
\begin{equation}\label{eq:pdagger_final}
p_i^\dagger(s,\nu)   = (2\tilde p_i  -1 )   p_i^*(s,\nu) - \tilde p_i +1. 
\end{equation}
Let $H_\nu$ be the event that the jump at time $\nu$ happened after the walker was trapped. Recall that, per~\eqref{eq:qi} we have $ \mathrm{P} \left\{   H_\nu \right\} = (1-p)^{|V_i|} = \tilde{q_i}$. Using again the law of total probabilities, 
\begin{multline}\label{eq:pitilde2}
\tilde{p}_i = \underbrace{\mathrm{P} \left\{   E'_\nu \vert  F_\nu \cap H_\nu \right\} }_{=0}       \mathrm{P} \left\{    H_\nu \vert F_\nu         \right\} \\
+ \mathrm{P} \left\{   E'_\nu \vert  F_\nu \cap  \overline{H_\nu} \right\}        \underbrace{\mathrm{P} \left\{    \overline{H_\nu} \vert F_\nu         \right\} }_{ = 1-\tilde{q_i}}. 
\end{multline} 
In the second term, 
\begin{equation}\label{eq:cld0}
	\mathrm{P} \left\{    E'_\nu \vert F_\nu \cap \overline{H_\nu}  \right\}  = \frac{\mathrm{P} \left\{    E'_\nu \cap F_\nu \cap \overline{H_\nu}  \right\} }{\mathrm{P} \left\{   F_\nu \cap \overline{H_\nu}  \right\} } 
\end{equation}
where the denominator is decomposed as 
\begin{multline}\label{eq:cld0denominator}
	\mathrm{P} \left\{   F_\nu \cap \overline{H_\nu}  \right\}  = 	\mathrm{P} \left\{    F_\nu \cap \overline{H_\nu}  \vert E'_\nu \right\}        \mathrm{P} \left\{  E'_\nu \right\} \\ 
	+ \mathrm{P} \left\{    F_\nu \cap \overline{H_\nu}  \vert \overline{E'_\nu} \right\}        \mathrm{P} \left\{  \overline{E'_\nu} \right\} 
\end{multline}
with $\mathrm{P} \left\{ E'_\nu  \right\} = p = 1- \mathrm{P} \left\{  \overline{E'_\nu} \right\}  $. Moreover, let $E^{(k)}_\nu $ be the event that $k$ out of $|V_i|-2$ out-neighbors of node $i$ are reachable at time $\nu$, so that
\begin{align}\label{eq:binomials1}
	&\mathrm{P} \left\{    F_\nu \cap \overline{H_\nu}  \vert E'_\nu \right\}      = \mathrm{P} \left\{  F_\nu \vert E'_\nu \right\}   \nonumber  \\
	&\qquad= \sum_{k=0}^{|V_i|-2} \mathrm{P} \left\{  F_\nu \vert E^{(k)}_\nu \cap E'_\nu \right\}   \mathrm{P} \left\{ E^{(k)}_\nu \vert E'_\nu  \right\}   \nonumber  \\
	& \qquad = \sum_{k=0}^{|V_i|-2} p \frac{1}{k+2} \times \binom{|V_i|-2}{k}  p^k (1-p)^{|V_i|-2-k}  \nonumber  \\
	& \qquad=\sum_{k=0}^{|V_i|-2}  \binom{|V_i|-2}{k}  \frac{1}{k+2}  p^{k+1} (1-p)^{|V_i|-2-k}. 
\end{align}
Using the same identity that allowed to obtain \eqref{eq:term1-corr}, 
\begin{equation}\label{eq:identitybinomials}
\sum_{k=1}^{n} \binom{n-1}{k-1}\frac{1}{k} p^k (1-p)^{n-k} = \frac{1-(1-p)^{n}}{n}, \quad n \geq 1,
\end{equation}
one eventually finds that the right-hand side of  \eqref{eq:binomials1} reads
\begin{equation}\label{eq:cld1}
\mathrm{P} \left\{    F_\nu \cap \overline{H_\nu}  \vert E'_\nu \right\}   = \frac{|V_i|p +(1-p)^{|V_i|-1}}{|V_i| (|V_i|-1)p}, \quad |V_i| \geq 2. 
\end{equation}
Similarly, for the remaining factor of \eqref{eq:cld0denominator} we have
\begin{align}\label{eq:binomials2}
	&\mathrm{P} \left\{    F_\nu \cap \overline{H_\nu}  \vert \overline{E'_\nu} \right\}    \nonumber   \\
	&\quad= \sum_{k=0}^{|V_i|-2} \mathrm{P} \left\{  F_\nu \cap \overline{H_\nu} \vert E^{(k)}_\nu \cap E'_\nu \right\}   \mathrm{P} \left\{ E^{(k)}_\nu \vert E'_\nu  \right\}  \nonumber  \\
	& \quad = \underbrace{\mathrm{P} \left\{ F_\nu \cap \overline{H_\nu} \vert E_\nu^{(0)} \cap \overline{E'_\nu}         \right\}}_{=p} \times \underbrace{\mathrm{P} \left\{   E_\nu^{(0)}  \vert \overline{E'_\nu} \right\}}_{=(1-p)^{|V_i|-2}}  \nonumber \\ 
	& \quad  +\sum_{k=1}^{|V_i|-2} \underbrace{ \mathrm{P} \left\{ F_\nu \cap \overline{H_\nu} \vert E_\nu^{(k)} \cap \overline{E'_\nu}         \right\} }_{=\frac{1}{k+1}p }   \, \times \! \! \!  \! \! \! \! \underbrace{\mathrm{P} \left\{   E_\nu^{(k)}  \vert \overline{E'_\nu} \right\}}_{= \binom{|V_i|-2}{k}p^k (1-p)^{|V_i|-2}}     \nonumber   \\
	& \quad=\sum_{k=0}^{|V_i|-2}  \binom{|V_i|-2}{k}  \frac{1}{k+1}  p^{k+1} (1-p)^{|V_i|-2-k}, 
\end{align}
and relying again on \eqref{eq:identitybinomials}, 
\begin{equation}\label{eq:cld2}
\mathrm{P} \left\{    F_\nu \cap \overline{H_\nu}  \vert \overline{E'_\nu} \right\}  = \frac{1-(1-p)^{|V_i|-1}}{|V_i|-1}, \quad |V_i| \geq 2. 
\end{equation}
Inserting \eqref{eq:cld1} and \eqref{eq:cld2} in \eqref{eq:cld0denominator} leads to writing \eqref{eq:cld0} as 
\begin{equation}\label{eq:cld0final}
\mathrm{P} \left\{    E'_\nu \vert F_\nu \cap \overline{H_\nu}  \right\}  = \frac{|V_i|p + (1-p)^{|V_i|}-1}{(1-|V_i|) \left(  (1-p)^{|V_i|}-1   \right)}, 
\end{equation} 
and eventually \eqref{eq:pitilde} becomes
\begin{equation}\label{eq:pitildefinal}
\tilde{p}_i = \frac{|V_i|p + (1-p)^{|V_i|}-1}{|V_i|-1}, \quad |V_i| \geq 2. 
\end{equation}
The expression of $p_i^\dagger (s,\nu)$ results from inserting \eqref{eq:pitildefinal} into \eqref{eq:pdagger_final}.

%\addcontentsline{toc}{section}{\refname}
\bibliography{CTRW_refs}

\end{document}